\documentclass{aa} 

\usepackage{graphicx}
\usepackage{url}
\usepackage{txfonts}
\usepackage{here}
\usepackage{threeparttable}
\usepackage[referable]{threeparttablex}
\usepackage{natbib}
\bibpunct{(}{)}{;}{a}{}{,}
\begin{document}

   \title{Spatial and Temporal Variations of the Chandra ACIS Particle-Induced Background and Development of a Spectral-Model Generation Tool}


\author{H. Suzuki \inst{1,2},
P. P. Plucinsky \inst{3},
T. J. Gaetz \inst{3} \and
A. Bamba \inst{1,4}
          }

   \institute{
   Department of Physics, The University of Tokyo, 7-3-1 Hongo, Bunkyo-ku, Tokyo 113-0033, Japan\\
\email{hiromasa050701@gmail.com}
\and Department of Physics, Faculty of Science and Engineering, Konan University, 8-9-1 Okamoto, Kobe, Hyogo 658-8501, Japan \\
\and Center for Astrophysics | Harvard \& Smithsonian, 60 Garden Street, Cambridge, MA 02138, USA \\
\and Research Center for the Early Universe, The University of Tokyo, 7-3-1 Hongo, Bunkyo-ku, Tokyo 113-0033, Japan
             }

   \date{Received **; accepted **}

\titlerunning{Particle-induced background of the {\it Chandra} ACIS}
\authorrunning{H. Suzuki et al.}

  \abstract
   {
In X-ray observations, estimation of the particle-induced background is important especially for faint and/or diffuse sources.
Although software exists to generate total (sky and detector) background data suitable for a given {\it Chandra} ACIS observation, no public software exists to model the particle-induced background separately.
   }
   {
We aim to understand the spatial and temporal variations of the particle-induced background of $Chandra$ ACIS obtained in the two data modes, VFAINT and FAINT.
We develop a tool to generate the particle-induced background spectral model for an arbitrary observation.
   }
   {
Observations performed with ACIS in the stowed position shielded from the sky and the {\it Chandra} Deep Field South data sets are used.
The spectra are modeled with a combination of the instrumental lines of Al, Si, Ni, and Au and continuum components.
The spatial variations of the spectral shapes are modeled by dividing each CCD into 32 regions in the CHIPY direction.
The temporal variations of the spectral shapes are modeled using all the individual ACIS-stowed observations.
   }
   {
Similar spectral-shape variations are found in VFAINT and FAINT data, which are mainly due to inappropriate correction of charge transfer inefficiency for events that convert in the frame-store regions as explained by \cite{bartalucci14}.
The temporal variation of the spectral hardness ratio is $\sim 10\%$ at maximum, which seems to be largely due to solar activity.
We model this variation by modifying the spectral hardnesses according to the total count rate.
Incorporating these properties, we have developed a tool {\tt mkacispback} to generate the particle-induced background spectral model corresponding to an arbitrary celestial observation.
As an example application, we use the background spectrum produced by the {\tt mkacispback} tool in an analysis of the Cosmic X-ray Background in the CDF-S observations. We find intensities of 3.10 (2.98--$3.21)\times 10^{-12}$ erg s$^{-1}$ cm$^{-2}$ deg$^{-2}$ in the 2--8 keV band and 8.35 (8.00--$8.70)\times 10^{-12}$ erg s$^{-1}$ cm$^{-2}$ deg$^{-2}$ in the 1--2  keV band, consistent with or lower than previous estimates.   
   }
   {
We model the spatial and temporal variations of the particle-induced background spectra of the $Chandra$ ACIS-I and the S1, S2, and S3 CCDs, and have developed a tool to generate a spectral model for an arbitrary celestial observation.
The tool {\tt mkacispback} is available at https://github.com/hiromasasuzuki/mkacispback.
   }

   \keywords{Instrumentation: detectors --
                Methods: data analysis --
                X-rays: general
               }

   \maketitle
%

\section{Introduction}\label{sec-intro}
For X-ray spectroscopy, background estimation is important especially for observations of faint and/or diffuse sources.
For point sources, the background can be estimated from nearby regions which are free from the source emission.
On the other hand, accurate background estimation from nearby regions is difficult for extended sources due to the contamination of the source emission and spatial variation of the background spectra.
The background consists of the cosmic background from Galactic and extragalactic sources (hereafter, "sky background") and the background induced by cosmic-ray particles (hereafter, "particle-induced background").
Both the observed sky and particle-induced backgrounds depend on the detector configuration and data reduction method.
There have been many efforts to study and model the particle-induced background for individual detectors onboard satellites (e.g., \citealt{tawa08} for {\it Suzaku} (XIS); \citealt{kuntz08, salvetti17, gastaldello17, marelli21} for {\it XMM-Newton} (EPIC); \citealt{wik14} for {\it NuSTAR}).
These works provided sufficient information on how one should model the background in an arbitrary observation.
The particle-induced background is complicated because many physical processes contribute to it, e.g., direct hit, generation of secondary particles, fluorescence line emissions, and radioactivation.
Thus, the particle-induced background in the X-ray energy range has been treated phenomenologically except for a few recent studies based on detector simulations (e.g., \citealt{hagino20} for the HXI onboard {\it Hitomi} and \citealt{lotti17, grant20} for the X-IFU and WFI onboard {\it Athena}).

In general, the particle-induced background depends on the satellite position (and possibly on attitude) and the solar activity.
In the case of {\it Suzaku}, the particle-induced background is thought to be free from temporal variation of solar activity due to its low-earth orbit.
The background spectra for an observation are thus determined by the satellite position and can be predicted from the earth-occultation data obtained at the same satellite location.
For {\it XMM-Newton}, because of its high altitude, the particle-induced background is highly affected by solar flares.
Temporal variations of the flux and spectral shape of the background were found. In addition, spatial variations of the instrumental fluorescence lines were also found.
The detector background spectra were able to be modeled by using the data from the parts of the CCD that were shielded from focussed X-rays from cosmic sources \citep{kuntz08}.

\begin{figure*}[htb!]
\centering
\includegraphics[width=12cm]{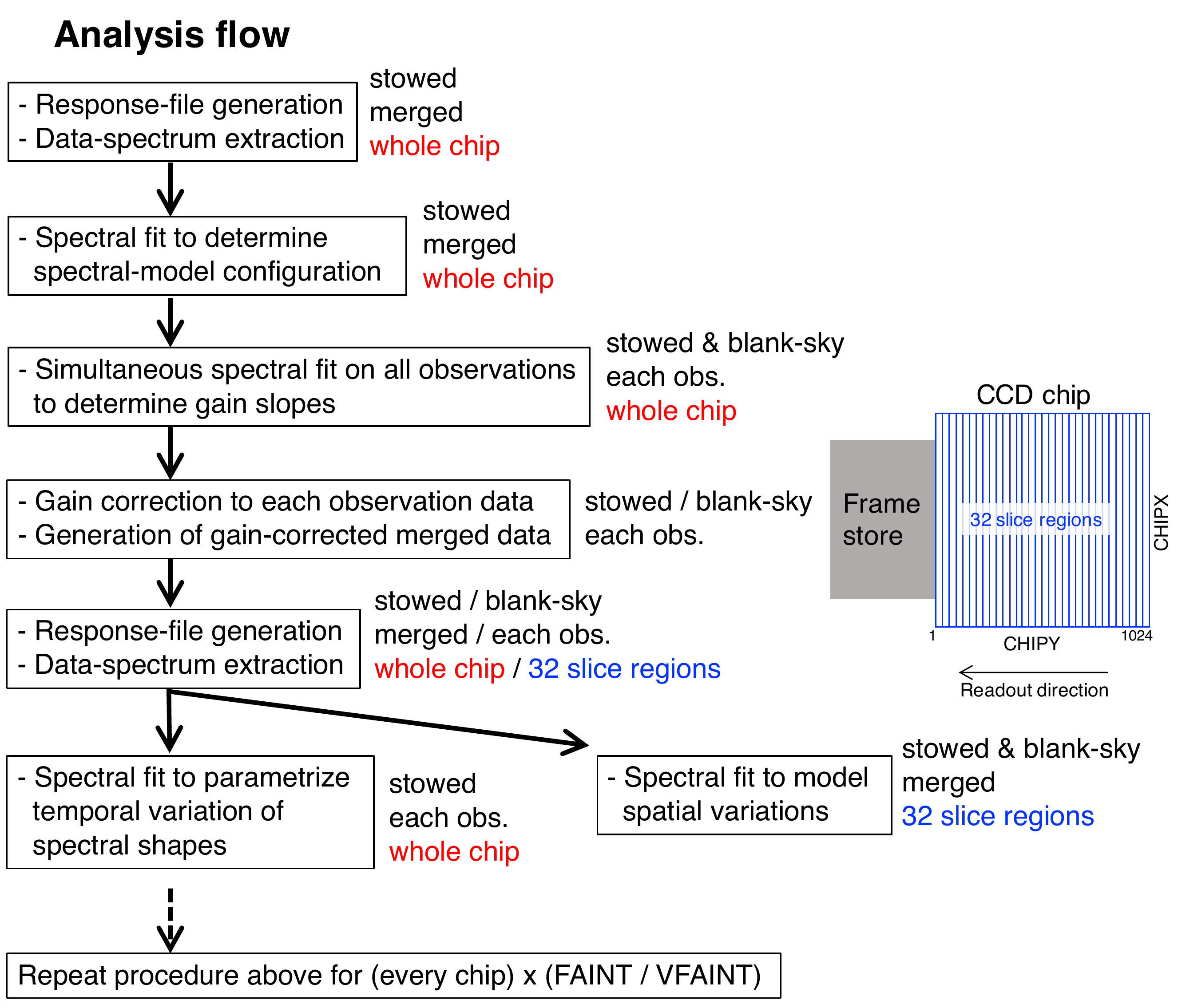}
\caption{Description of our analysis flow.
The ``stowed'' / ``blank-sky'', ``merged'' / ``each obs.'', and ``whole chip'' / ``32 slice regions'' stand for the observation types, data types, and detector regions for which the analysis processes are applied, respectively.
After the gain correction part, the ``merged'' and ``each obs.'' mean the gain-corrected-and-merged data and each of the gain-corrected observation data.
A schematic drawing of a CCD chip with the 32 slice regions along CHIPY, readout direction, and frame-store region is also shown.
\label{fig-anaflow}}
\end{figure*}

In the case of {\it Chandra}, most observers can extract a useful background region that is free from emission from the source given {\it Chandra}'s superb imaging capabilities. However, there are sources that are large enough such that they fill the entire {\it Chandra} field-of-view and such an approach is not possible. For these cases, the  {\it Chandra} X-ray Center (CXC) has made available the CALDB ``blank-sky'' data sets and software to create a background events lists suitable for the observation of interest\footnote{\url{https://cxc.harvard.edu/ciao/ahelp/blanksky.html}}. These blank-sky data sets include the sky and particle-induced background components as they are derived from {\it Chandra} pointings with point sources removed from various locations on the sky \citep{markevitch03}. Two disadvantages of this approach are that it combines the sky and particle-induced background components and it averages the sky background from different directions. For some applications it would be advantageous to model the sky and particle-induced background components separately. This would require a model of the particle-induced background of the detector.
\cite{bartalucci14} studied the spatial and temporal variation of the particle-induced background of the ACIS-I CCDs in the very faint (VFAINT) mode.
They parameterized the spectra with multiple line components, a power-law and an exponential function.
The spatial variation was found to be largely due to the ``frame-store lines'' (``daughter lines'' in \citealt{bartalucci14}), which are the emission lines detected in the frame-store regions of the ACIS array during frame readout (see, e.g., Fig.~\ref{fig-anaflow} for the position of the frame-store region). This variation was seen only along the readout direction as expected.
The temporal variation, i.e., the short-term variation depending on the satellite position and long-term variation due to the solar activity, was able to be described by changing the normalizations of the spectra without changing their shapes.

This work aims to characterize the spatial and temporal variations of the particle-induced background spectra of the {\it Chandra} ACIS-I and S1, S2, and S3 CCDs obtained in both the FAINT and VFAINT modes, and to develop a spectral-model generation tool for practical uses.
The data obtained in FAINT mode include higher particle-induced background contributions than those in VFAINT mode.
Also, the back-illuminated (BI) CCDs have a higher event rate in the accepted grade set (g02346) due to particles than the FI CCDs owing to the fact that more of the particle events appear in the rejected grade set for the FI CCDs.
These properties are suitable for a detailed study to investigate their spatial and temporal variations.
Throughout the paper, errors indicate a $1~\sigma$ confidence range.

\begin{figure}[htb!]
\centering
\includegraphics[width=8cm]{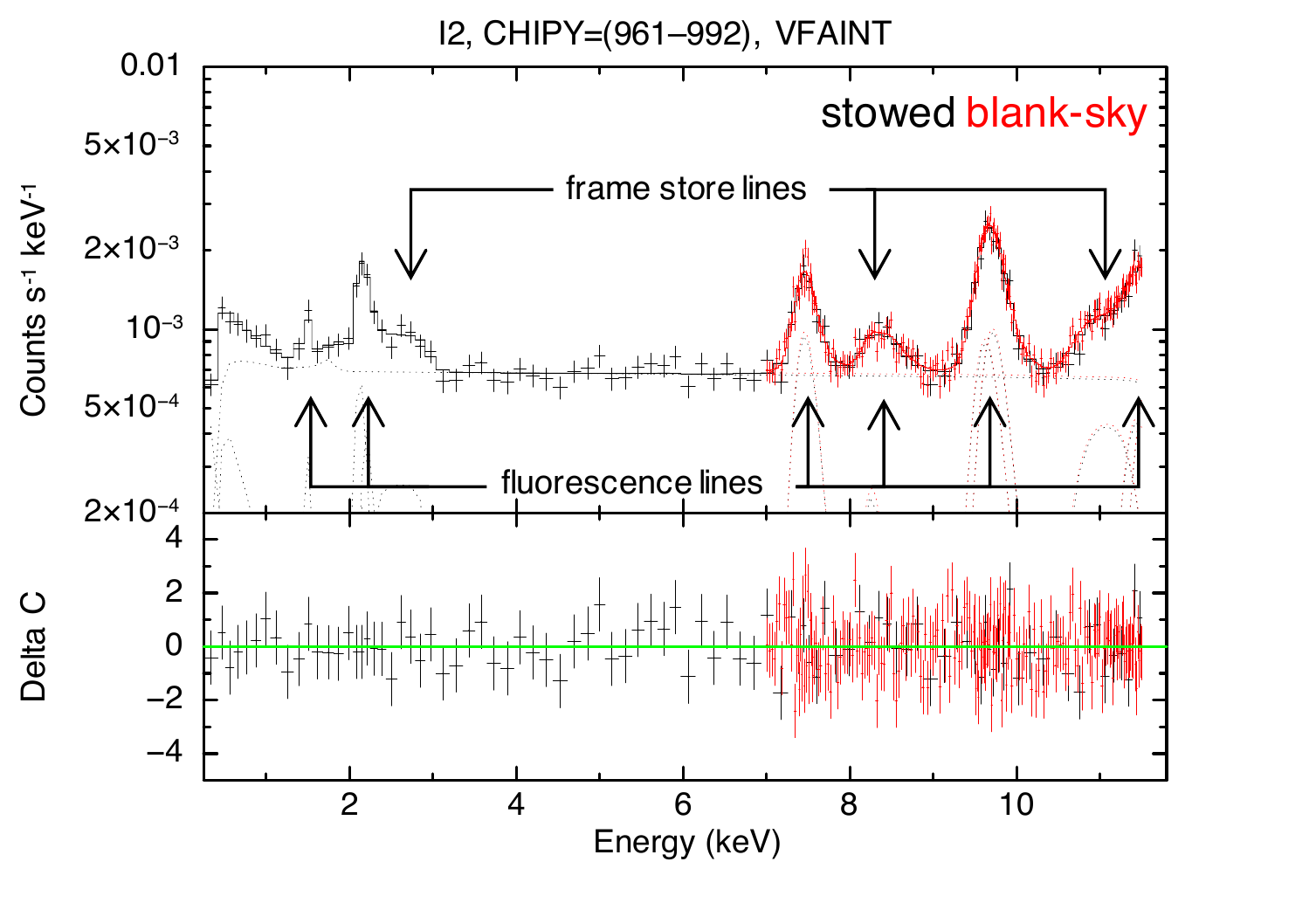}
\caption{An example spectrum of the observational particle-induced background extracted from a small region $({\rm CHIPX}, {\rm CHIPY}) = (1 \colon 1024 ,961 \colon 992)$ of the I2 CCD.
The data are extracted from the merged ACIS-stowed (black) and CDF-S blank-sky (red) observations.
The upper and lower panels present the fluxes and residuals, respectively.
The data and the best-fit models are shown with the crosses and solid lines.
The model components are presented with the dotted lines.
We note that the spectral extraction region is far from the frame-store region so that the CTI effects are important.
\label{fig-example}}
\end{figure}

\section{Data Reduction and Analysis}\label{sec-analysis}
We use two types of data sets to obtain spectral models to describe the particle-induced background for ACIS: the ``ACIS-stowed'' observations and the {\it Chandra} Deep Field South observations (hereafter, ``CDF-S blank-sky'' observations).
The ACIS-stowed observations are conducted with ACIS out of the focal position of the telescope, so that the events originate only from the particle-induced background.
The CDF-S blank-sky data sets consist of the sky background (unresolved Galactic and extragalactic sources) and particle-induced background.
We note that this blank-sky data set is different from the CXC's CALDB blank-sky data sets, which are used only for verification of our background modeling in Section~\ref{sec-verification}.
We use the {\it Chandra} Deep Field South observations because they have longer exposures so that the statistics are better.
For the CDF-S blank-sky observations, we do not remove point sources in the field of view.
Because we only use the CDF-S blank-sky data for the energies above 7~keV, the point sources contribute less than a few percent of the total counts \citep{bartalucci14}.
The observation logs for each of these two data types are summarized in Tables~\ref{tab-stowed} and \ref{tab-blanksky}.
The ACIS-stowed observations range from 2002 to 2016, with a total exposure of $\sim1$~Ms.
The CDF-S blank-sky observations range from 2007 to 2016, with a total exposure of $\sim7$~Ms.
For each observation, the data are reprocessed to create a ``level=2'' event file. In the analysis below, spectra and corresponding response files are created based on the level=2 files.
The ``merged'' events files for each of the ACIS-stowed and CDF-S blank-sky data sets are generated as well by merging all the observations listed in Tables~\ref{tab-stowed} and \ref{tab-blanksky}, respectively.

In the data reduction, we use CIAO (v4.11; \citealt{fruscione06}) and HEAsoft (v6.20; \citealt{heasarc14}).
In the spectral analysis, we use XSPEC (v12.9.1; \citealp{arnaud96}).
The {\it C}-statistic, which has been implemented in XSPEC as {\it cstat}, is used in all the spectral fitting \citep{cash79, kaastra17}.

The overall analysis procedure is summarized in Fig.~\ref{fig-anaflow}.
Our approach for the spectral modeling is based on \cite{bartalucci14} and the presentation by T. Gaetz at the 14th IACHEC 2019\footnote{\url{https://iachec.org/wp-content/presentations/2019/WGV_Gaetz.pdf}}.
As the first step, the energy spectra and response matrices are generated for individual CCD chips for the ACIS-stowed observations.
Using these, we generate a spectral model to describe the data.
The resultant spectral model is a combination of multiple Gaussian lines and several continuum models (power-law, broken power-law, and exponential function).
The line components are composed of Al, Si, Au, and Ni lines. The line centroids are fixed to the literature values \citep{bearden67} shown in Table~\ref{tab-lines}.
As described in \cite{bartalucci14}, the spectra include broad line components produced by the inappropriate correction of charge transfer inefficiency (CTI) for events that convert in the frame-store regions of the CCDs (frame-store lines).
The spectral shape of the frame-store line can be approximated by a function
\begin{equation}\label{eq-fsline}
\begin{aligned}
C \quad &(E_{\rm min} < E < E_{\rm max})\\
0 \quad &\text{(else)},
\end{aligned}
\end{equation}
where the $C$ is a constant, and $E_{\rm min}$ and $E_{\rm max}$ determine the boundaries of the component.
Fig.~\ref{fig-example} shows an example of the merged ACIS-stowed and CDF-S blank-sky spectra extracted from a small region of the I0 CCD.
The broad line components seen at $\approx 2.7$~keV and $\approx 10.7$~keV are the frame-store lines originating from Au-M and Au-L lines, respectively\footnote{These components have smooth shapes due to the CCD's finite energy resolution.}.
All the parameters of the model components other than the line centroids are treated as free parameters in our analysis.

We note two things about our analysis.
First, in the ACIS-stowed observations, only one observation (OBSID: 62678) for the I1 CCD is available.
Thus, the spatial and temporal variations of the I1 data are substituted by those of the I0 data, given the expected similar properties\footnote{\url{https://cxc.cfa.harvard.edu/contrib/maxim/stowed/i01/}}.
This treatment is the same as that taken in \cite{bartalucci14}.
Second, the CDF-S blank-sky data are not available for the S1 and S3 CCDs.
Although we find that spectral models for the S1 and S3 CCDs can be obtained only with the ACIS-stowed data, we also use the CALDB blank-sky data sets, which include the S1 and S3 CCD data, to verify these spectral models (Section~\ref{sec-verification}).

\begin{table}[htb!]
\centering
\caption{Detector line properties included in this work \citep{bearden67}.
\label{tab-lines}}
\begin{threeparttable}
\begin{tabular}{l l l l}
\hline\hline
Element  &  Type  &  Energy (keV)  & Frame-store line\tnote{a}  \\\hline
Al  &  K$\alpha$  &  1.487  & Y \\
&  K$\beta$  &  1.557  & Y \\ 
Si  & K$\alpha$  &  1.740  & N\\
Au  &  M$\alpha_1$  &  2.123  & Y \\
&  M$\alpha_2$  &  2.118  & Y\\
&  M$\beta$  &  2.205  & Y\\
&  M$\gamma$  &  2.410  & Y\\
Ni  &  K$\alpha_1$  & 7.478  & N\\
&  K$\alpha_2$  & 7.461 & N\\
&  K$\beta$  &  8.265  & N\\
Au  &  L$_1$  &  8.494  & Y \\
&  L$\alpha_1$  &  9.713  & Y \\
&  L$\alpha_2$  &  9.628  & Y \\
&  L$\beta_1$  &  11.442  & Y \\
&  L$\beta_2$  &  11.585  & Y \\
\hline
\end{tabular}
\begin{tablenotes}
\item[a] The ``Y'' (``N'') indicates that the frame-store lines are included (not included because they are faint) in our spectral modeling.
\end{tablenotes}
\end{threeparttable}
\end{table}

\begin{figure*}[htb!]
\centering
\includegraphics[width=16cm]{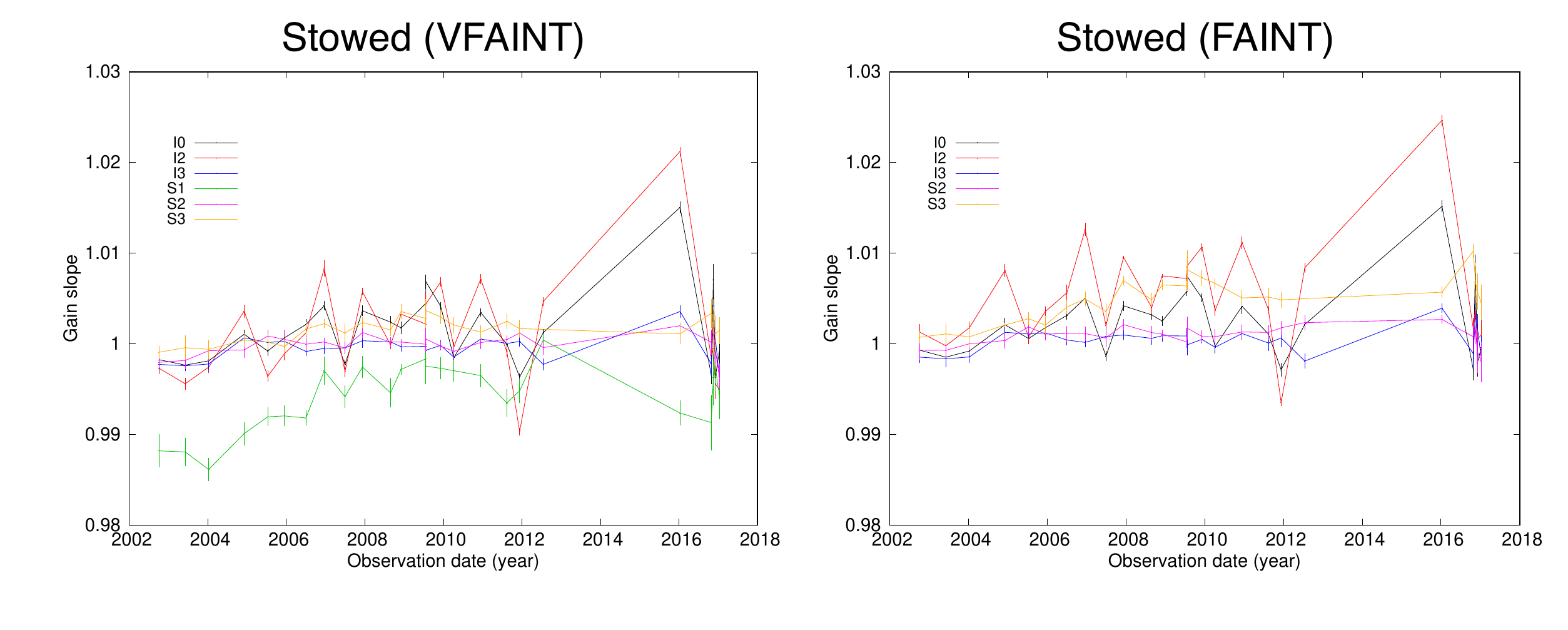}
\caption{Gain slopes for individual ACIS-stowed observation data.
The S1 data obtained in FAINT mode is excluded because the gain slopes are not determined well in this case due to less prominent line emissions.
\label{fig-gains-stowed}}
\end{figure*}

\begin{figure*}[htb!]
\centering
\includegraphics[width=16cm]{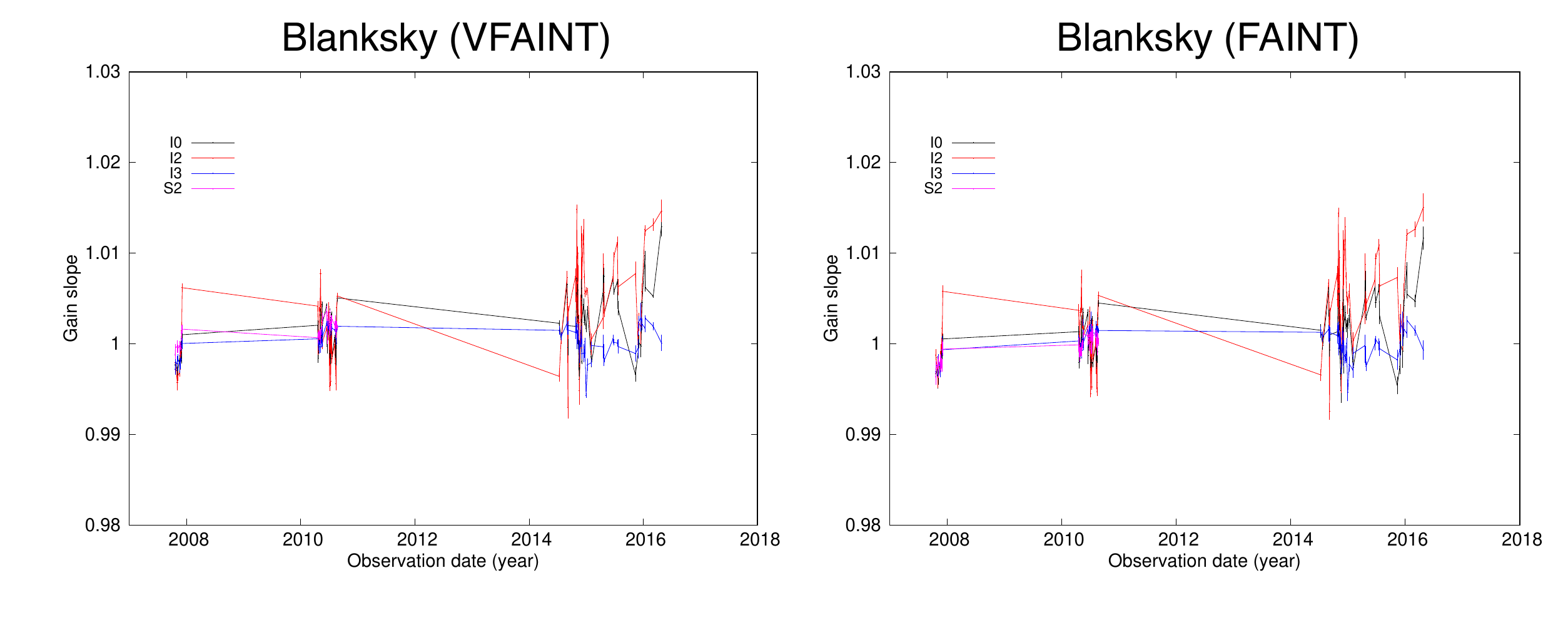}
\caption{Gain slopes for individual CDF-S blank-sky observation data.
\label{fig-gains-blanksky}}
\end{figure*}

\subsection{Detector gain correction}
In principle, the detector gains are well-calibrated in the level=2 events files at energies between 1.5 and 6.0~keV given that these are the energies of the strongest lines in the calibration source onboard.
At energies above 6.0~keV, the gain calibration may be less accurate and may lead to residuals around the strong lines of Ni at $\approx 7.5$~keV and Au at $\approx9.7$~keV.
In order to investigate the gain variations, we fit the spectra extracted from individual observations simultaneously with the model configuration obtained above.
The energy ranges of 0.25--11.5~keV for the ACIS-stowed and 7.0--11.5~keV for the CDF-S blank-sky spectra are used.
The spectra are extracted from the entire CCDs.
The energy range for the CDF-S blank-sky observations are selected based on \cite{bartalucci14} to avoid contamination of the sky background.
We fit the data with free gain offset and slope parameters, but find that the gain offset values are small (scatter by a few eV, typically, and are mostly consistent with zero).
Thus, for the sake of the reduction of computational costs, only their slopes are treated as free parameters.
An offset of zero and a small deviation from a slope of 1.0 is consistent with an accurate gain calibration below 6.0~keV and a small adjustment at energies above 6.0~keV.
The gain slopes are constrained by the energy centroids of the line emissions, particularly by those of Ni-K$\alpha$ and Au-L$\alpha$ emissions.

The resultant gain slope values versus observation date are presented in Fig.~\ref{fig-gains-stowed} and Fig.~\ref{fig-gains-blanksky}.
The gain values for VFAINT and FAINT modes are roughly consistent with each other.
In some observations, slight inconsistencies between them are seen. These are probably due to high continua and thus less prominent line emissions in FAINT mode, which may lead to a less accurate estimation of the gain slopes.
Generally, this level of the discrepancies will not affect the spectral modeling, but it might affect some cases as discussed in Section~\ref{sec-verification}.
The I3 and S2 CCDs are found to show relatively stable gain slopes with time, whereas those of the I0, I2 and S1 CCDs vary greatly within $\sim\pm \, 2 \%$.
We apply the gain correction to each observation based on the gain slopes obtained above, and generate gain-corrected spectra and response matrices.
Then, the gain-corrected-and-merged spectra are generated as well.
For the S1 CCD in FAINT mode, the gains cannot be determined because of the particularly high continuum, so that no gain correction is applied to it.

\begin{figure*}[htb!]
\centering
\includegraphics[width=12cm]{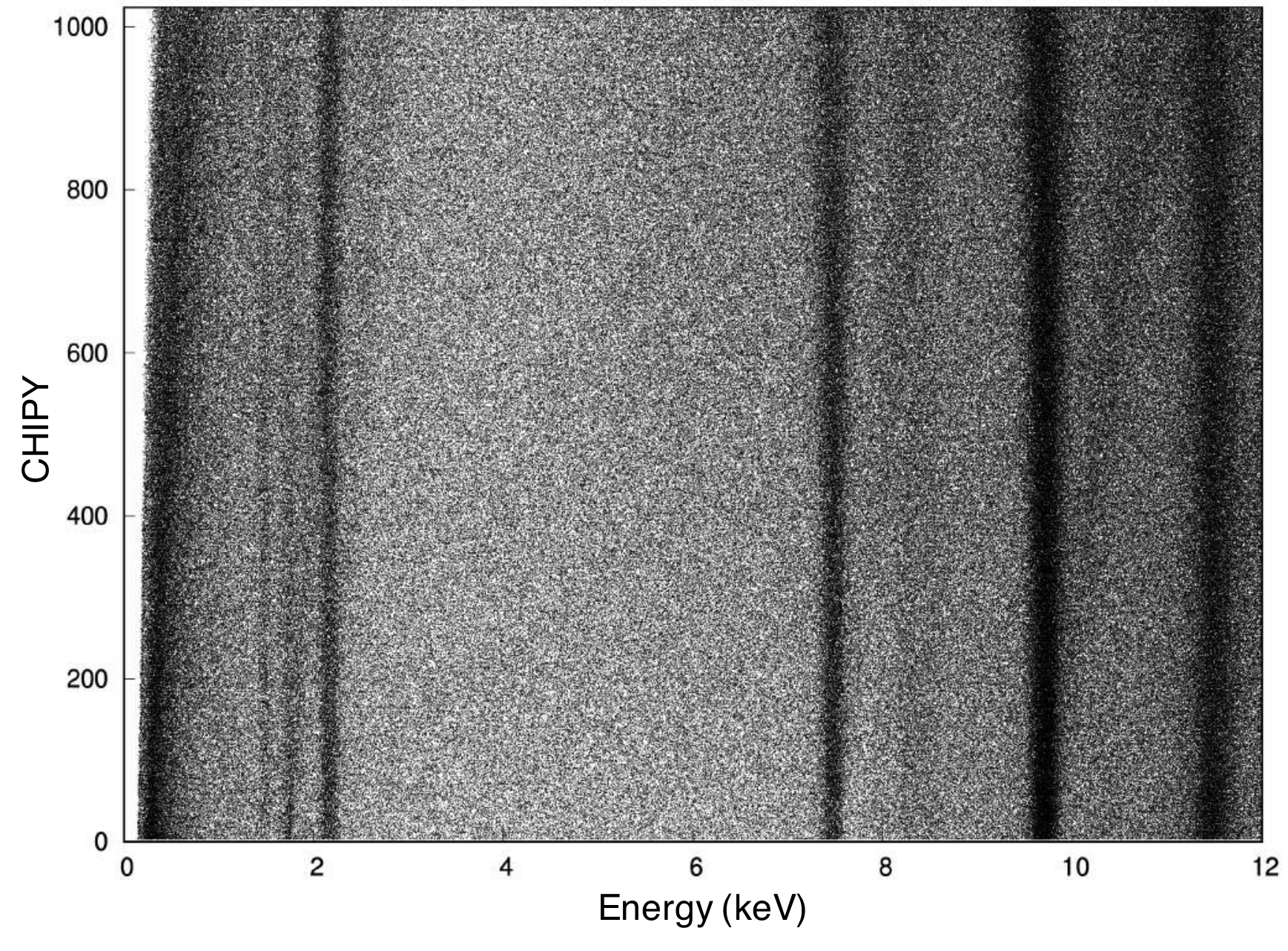}
\caption{Scatter plot of the particle-induced background events on the CHIPY-energy plane extracted from the merged ACIS-stowed data.
This is the sum of the I0, I2 and I3 events. Darker regions include larger numbers of events.
\label{fig-chipy-energy}}
\end{figure*}

\subsection{Spatial variation of the spectra}
In Fig.~\ref{fig-chipy-energy}, we present CHIPY-energy scatter plots for stowed events extracted from I0, I2 and I3 showing various X-ray fluorescence lines. 
The vertical features are fluorescent lines (Al-K, Si-K, Au-L, Ni-K, and Au-K) which convert in the imaging part of the array.
The CHIPY-dependent CTI correction has been applied which corrects for the decrease in Pulse Hight Amplitude (PHA) due to CTI. 
Many of these lines also have the frame-store lines, which are fainter lines associated with reported energies increasing with CHIPY.
In processing the data, the software applies an energy- and CHIPY-dependent correction to the event PHAs.
Because the frame store is undamaged by the radiation dose the FI chips experienced early in the mission, the CTI in the frame store is close to zero.
The frame-store events experience effectively no CTI, so the application of the CTI correction results in an inappropriate increase in the frame-store event PHAs, giving an approximately linear variation of the frame-store line energies.

\begin{figure*}[htb!]
\centering
\includegraphics[width=16cm]{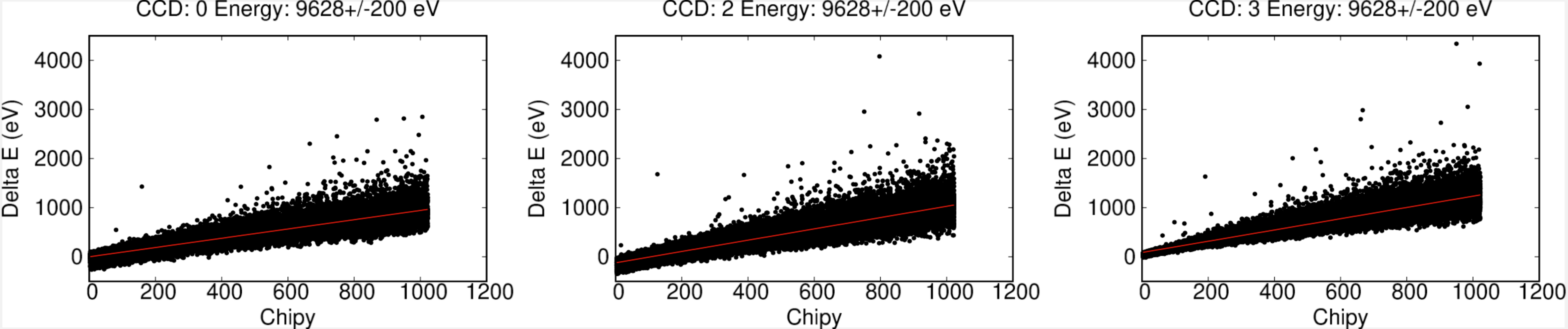}
\caption{Scatter plots of the events on the CHIPY-$\Delta E$ plane for I0 (CCD0), I2 (CCD2) and I3 (CCD3).
The events are extracted from the gain-corrected-and-merged ACIS-stowed data in VFAINT mode.
The extraction energy range is $9.628 \pm 0.2$~keV.
The red sold lines represent the best-fit linear functions.
\label{fig-deltaE}}
\end{figure*}

We investigate the CHIPY dependence of the frame-store line energies using an approach similar to that of \cite{bartalucci14}, in order to determine the energy bounds of the frame-store lines.
The event list reports for each event a ``PHA\_RO'' (readout PHA, depending on the charge collected in the $3 \times 3$ event detection island), and ``PHA'' (the result after applying CTI correction).  The ``ENERGY'' column provides the event energies, the result of adding a time-dependent gain (``TGAIN'') correction\footnote{\url{https://cxc.cfa.harvard.edu/ciao/why/acistgain.html}}, and using the DETGAIN information.
For clarity, we refer to the PHA\_RO values as $\mathrm{pha}_\mathrm{raw}$, the CTI-corrected PHA values as $\mathrm{pha}_\mathrm{crt}$, and the ENERGY values as $E$.
One can approximate the energy displacement of the frame-store events as
\begin{equation}
\Delta E = \frac{\mathrm{pha}_\mathrm{crt} - \mathrm{pha}_\mathrm{raw}}{\mathrm{pha}_\mathrm{raw}} E.
\end{equation}
The CHIPY-$\Delta E$ plots for individual fluorescence frame-store lines show linear-like correlations with larger $\Delta E$ values at larger CHIPY values.
An example is shown in Fig.~\ref{fig-deltaE}, which shows the CHIPY versus $\Delta E$ variation for the Au-L$\alpha_2$ line for the CCDs I0, I2, and I3.
The red lines are linear fits for $\Delta E$ versus CHIPY plots.
The I2 chip has a larger TGAIN correction, resulting in the fit intercepting the CHIPY=0 axis at a visibly negative value for energy; all of the chips have negative offsets, but these are usually small.
In principle, we can calculate the linear trend for $\Delta E$ and the energy bounds of the frame-store lines from the CHIPY-$\Delta E$ plots, which can be extracted from any (sufficiently long) celestial observation of interest.
However, we find no significant variations in the CHIPY-$\Delta E$ plots among observations and we can describe the spectra extracted from the whole-chip regions of any observation by an average model determined by the merged ACIS-stowed data (see Sec.~\ref{sec-verification} and Figs.~\ref{fig-vf-i0-eachobs}--\ref{fig-f-s3-eachobs})\footnote{The slope in the CHIPY-$\Delta E$ plot varies from CCD to CCD.}.
As for spectral variations along CHIPX, only small (mostly less than a few percent) variations of the spectral shapes are found for all the ACIS-I and S1, S2, and S3 CCDs (as partly noted by \citealt{bartalucci14}), based on our analysis of the merged ACIS-stowed data.
We note that the spatial variations of the FAINT data are similar to those of the VFAINT data -- the differences mostly appear in the spectral continuum shapes.

\begin{figure*}[htb!]
\centering
\includegraphics[width=16cm]{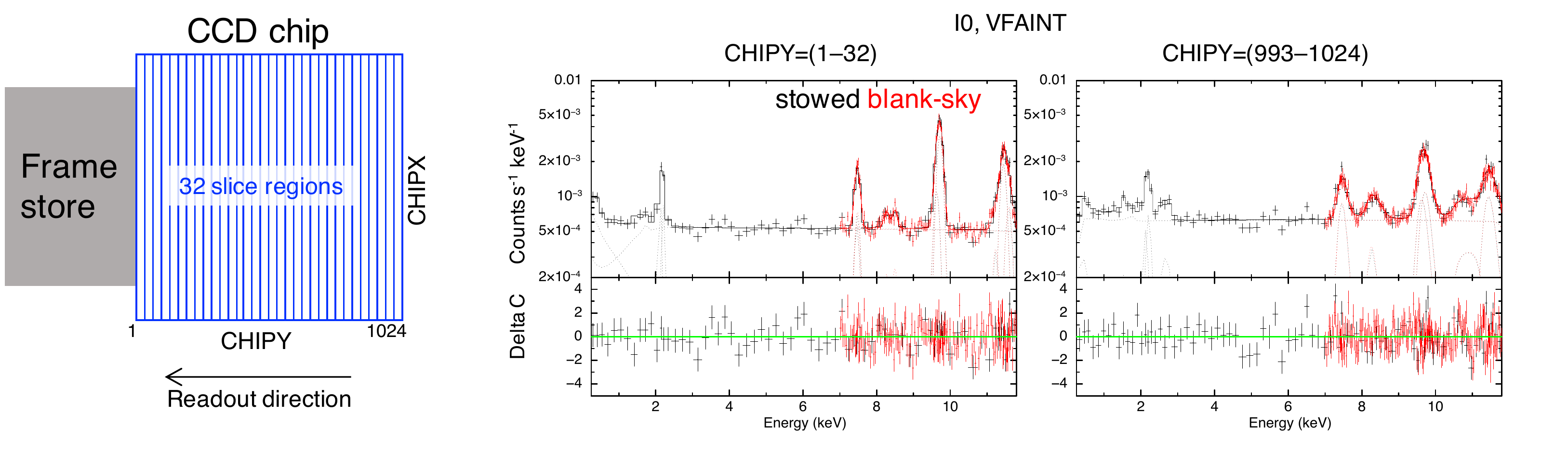}
\caption{An example of the spatial variation of the particle-induced background spectra.
Two spectra extracted from the low- and high-CHIPY regions of the merged ACIS-stowed data (I0, VFAINT) are presented.
The left panel shows the 32 slice regions along CHIPY.  
In each of the two right panels, the upper panel presents the data and model fluxes, whereas the lower presents their residuals.
The black and red crosses represent the ACIS-stowed and CDF-S blank-sky observations, respectively.
The solid and dotted lines are the total models and their components, respectively.
\label{fig-spatial}}
\end{figure*}

We extract the data and response matrices for 32 slice regions along the CHIPY axis (see Fig.~\ref{fig-spatial}).
These regions are defined as $({\rm CHIPX}, {\rm CHIPY}) = (1\colon1024, 1\colon32), (1\colon1024, 33\colon64), ..., (1\colon1024, 993\colon1024)$.
In order to model the spatial variations, we simultaneously fit the gain-corrected-and-merged ACIS-stowed and CDF-S blank-sky spectra extracted from each slice region.
As a validation of the fits, we confirm here that each fitting result yields $C$-stat/d.o.f.$ < 1.1$ (d.o.f. is $\approx 800$--1100), where d.o.f. stands for degree of freedom.
We repeat this process for the 32 regions for all CCDs used and both observation modes.
An example of the spectral differences along CHIPY is shown in Fig.~\ref{fig-spatial}.
Except for the difference in the detector responses, the largest difference is in the energy centroids and strengths of the frame-store lines.
In this process, we treat all of the spectral parameters (including the energy bounds of the frame-store lines) other than the detector line centroids as free parameters.
We obtain 32 sets of the model parameters for each CCD and each observation mode.
These are the base models for the spectral-model generation tool described in Sec.~\ref{sec-tool}.

\begin{figure*}[htb!]
\centering
\includegraphics[width=16cm]{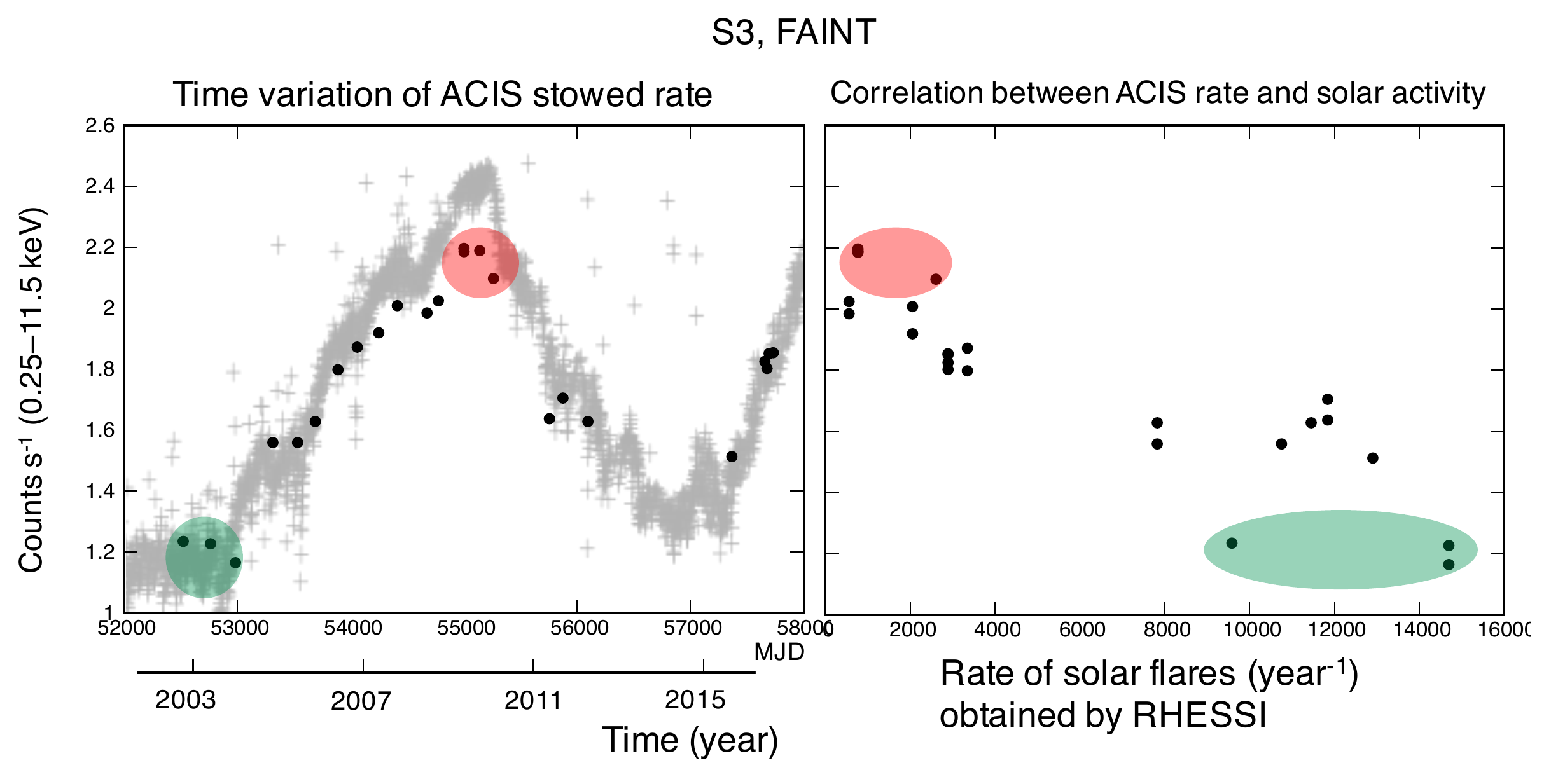}
\caption{Temporal variation of the particle-induced background rate of ACIS-S3, FAINT mode.
The transparent crosses are based on the rate of S3 events rejected onboard for exceeding the upper-level PHA threshold; they are scaled by $\approx 2\%$ to be compared to the particle-induced background rate.
The red and green ellipses enclose the observations at especially high and low particle-induced background rates.
The ``MJD'' stands for the Modified Julian Date.
\label{fig-time}}
\end{figure*}

\begin{figure*}[htb!]
\centering
\includegraphics[width=16cm]{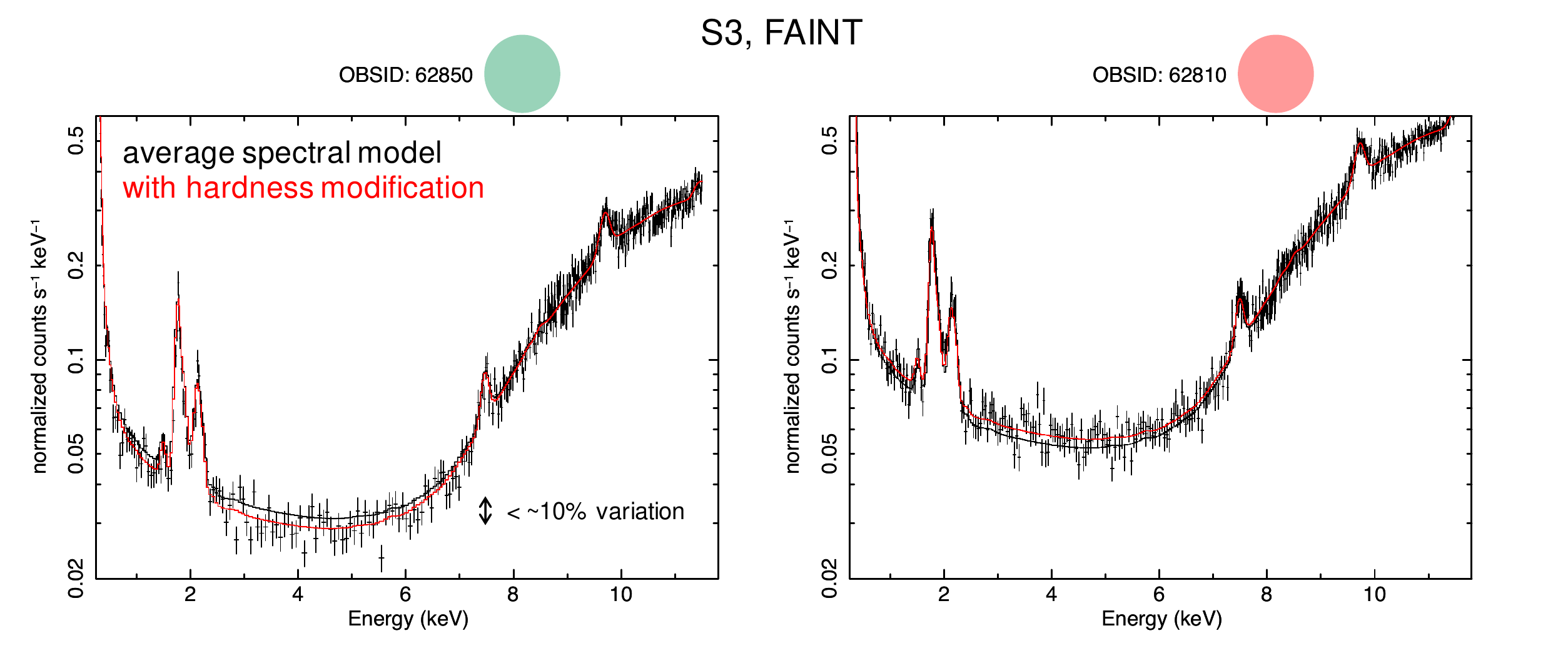}
\caption{Example of temporal variation of the particle-induced background spectral shapes.
The black crosses represent the data. The black and red solid lines show the average spectral models and the models with the optimized hardness modification.
We note that the OBSID 62850 and 62810 show the lowest and highest count rates in the ACIS-stowed observations, respectively.
\label{fig-time2}}
\end{figure*}

\subsection{Temporal variation of the flux and spectral shape}
In this section, we investigate the temporal variations of the particle-induced background spectra.
Fig.~\ref{fig-time} shows the long-term variation of the particle-induced background rate and its correlation with the solar activity (solar-flare rate)\footnote{\url{https://hesperia.gsfc.nasa.gov/rhessi3/data-access/rhessi-data/flare-list}}.
The left panel of Fig.~\ref{fig-time} compares the temporal variation of the ACIS-stowed S3 count rate (0.25--11.5~keV) to an estimate of the total S3 cosmic-ray rate\footnote{\url{https://space.mit.edu/~cgrant/cti/cti120.html}}.
The S3 cosmic-ray rate is based on tallies of the ``events'' which exceed the S3 upper event PHA threshold (corresponding to event energies exceeding $\approx 12$~keV).
In the right panel of Fig.~\ref{fig-time}, a negative correlation between the particle-induced background and the solar activity is seen.
This indicates that the temporal variation of the particle-induced background is largely due to solar activity, and thus, the incoming cosmic-ray flux\footnote{Since cosmic-ray particles are decelerated by solar wind in the solar system, the incoming cosmic-ray flux anti-correlates with solar activity (e.g.,~\citealt{mizuno04}).}.

\begin{figure*}[htb!]
\centering
\includegraphics[width=16cm]{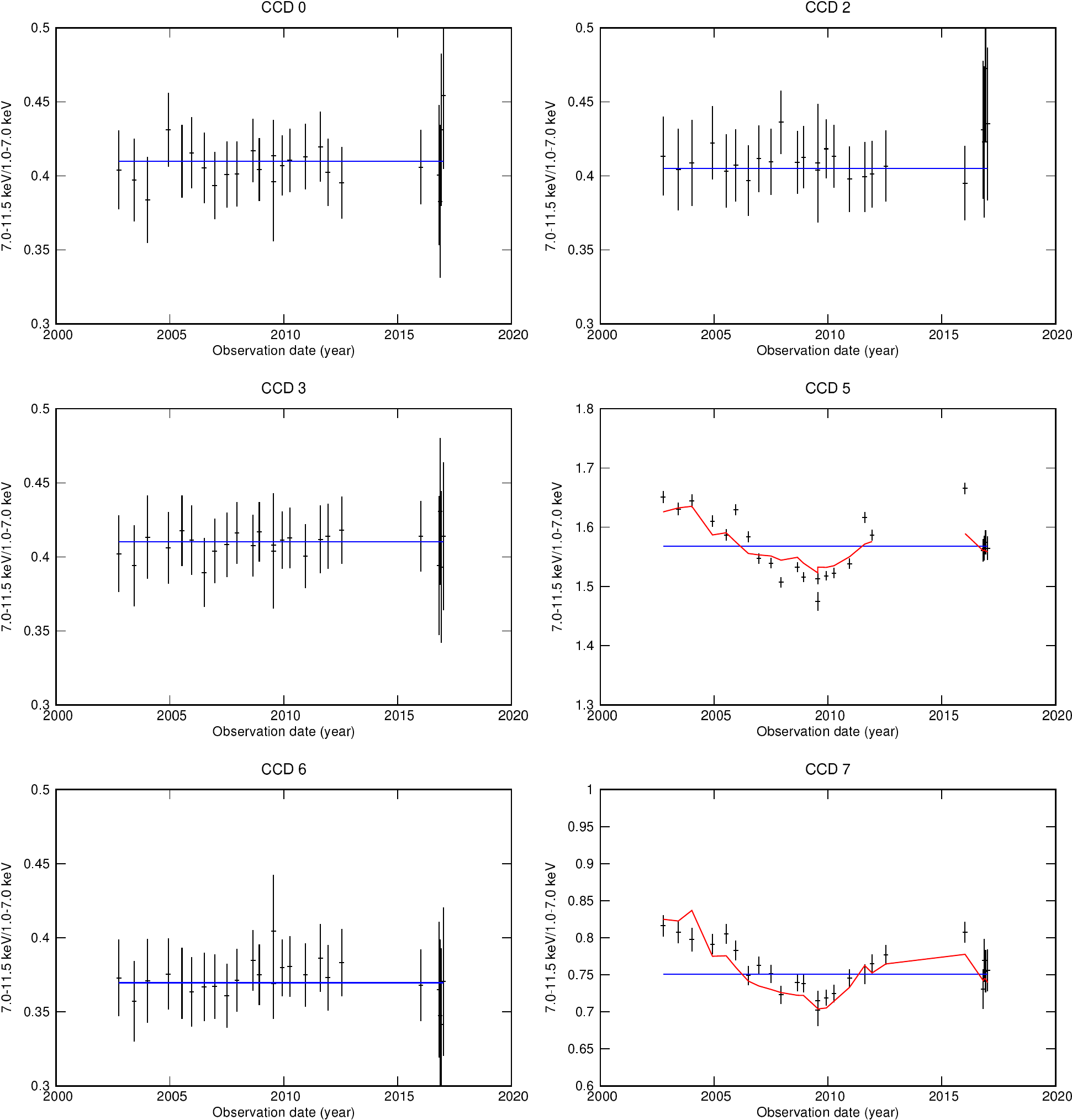}
\caption{Hardness ratio (7.0--9.0~keV/1.0--7.0~keV) versus observation date.
The black crosses and blue and red lines represent the data, average models, and models with the optimized hardness modification, respectively.
The data and models are the ACIS-stowed observations processed in VFAINT mode.
CCD 0, 2, 3, and 5--7 correspond to I0, I2, I3 and S1--S3, respectively.
\label{fig-hardness-vf}}
\end{figure*}

\begin{figure*}[htb!]
\centering
\includegraphics[width=16cm]{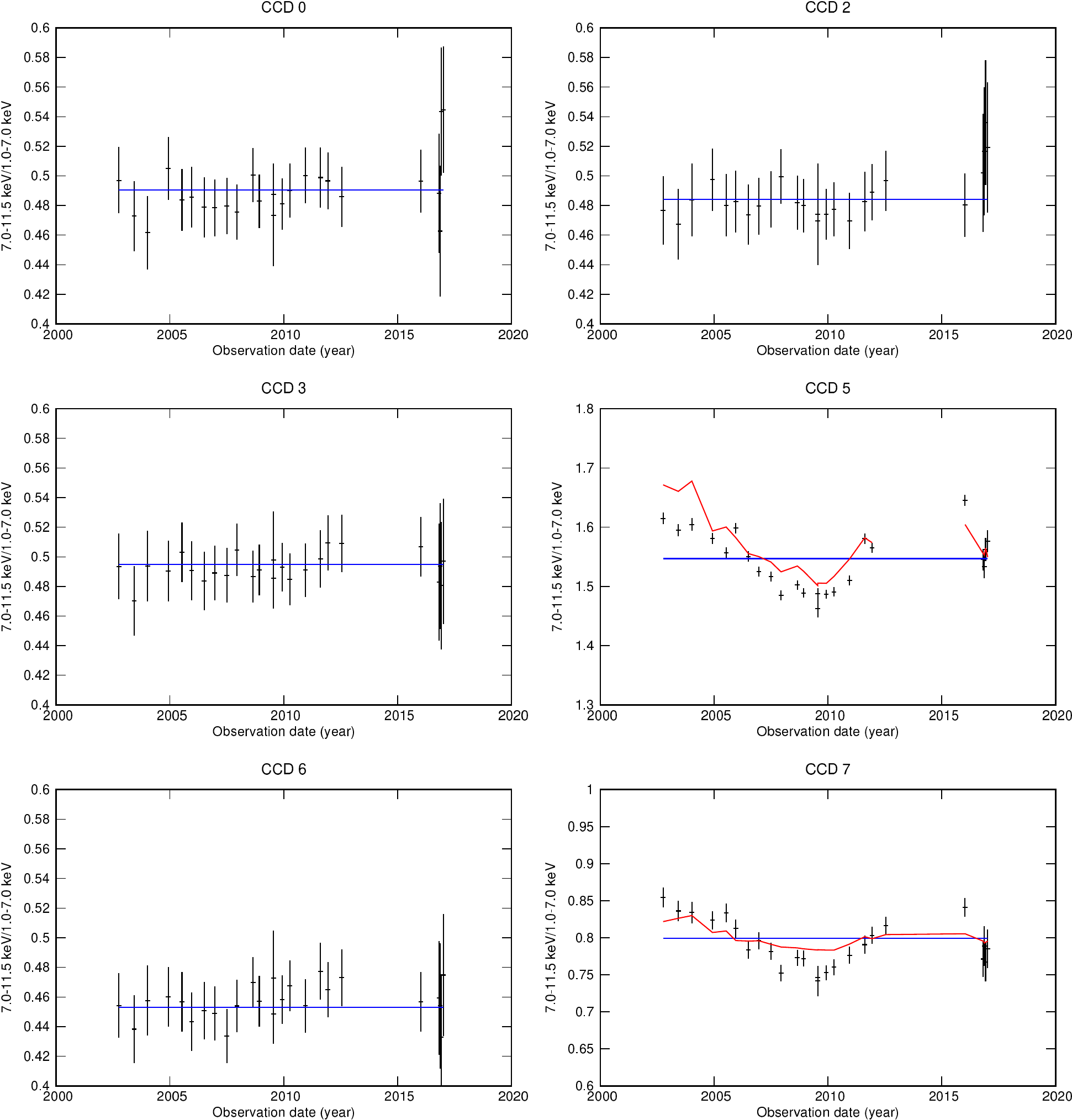}
\caption{Same as Fig.~\ref{fig-hardness-vf}, but for the data processed in FAINT mode.
\label{fig-hardness-f}}
\end{figure*}

Comparing the spectral shapes of the data to the average models which are obtained by summing the 32 sets of the base models (corresponding to the 32 $\Delta$CHIPY regions), we find that the continuum shape varies and thus the hardness ratio also varies, as noted by \cite{bartalucci14}.
Fig.~\ref{fig-time2} exhibits the variation of the spectral shapes by comparing the data to our average models for two extreme observations, OBSIDs 62850 and 62810.
OBSID 62850 had one of the lowest overall background rates and OBSID 62810 had one of the highest.
Discrepancies between the spectral normalizations of the data and average models for $\sim 1.0$--7.0~keV can be seen in Fig.~\ref{fig-time2}.
Figs.~\ref{fig-hardness-vf} and \ref{fig-hardness-f} show the temporal variation of the hardness ratio (7.0--9.0~keV/1.0--7.0~keV) for the ACIS-stowed observations in VFAINT and FAINT modes, respectively.
The average models obtained above are plotted in Figs.~\ref{fig-hardness-vf} and \ref{fig-hardness-f} for comparison.
The two BI CCDs show significant variation of the hardness ratio of $\lesssim \pm 10\%$.
The FI CCDs show no significant variations in spectral shape.
The tendency of the hardness-ratio variation of the BI CCDs with time is similar to that of the total particle-induced background rate (see Fig.~\ref{fig-time}), implying that the cause of these shape variations is also related to the cosmic-ray flux.

Here we model the hardness-ratio variations only for the BI CCDs.
As a simple model of this temporal variation of the spectral shapes, we let the continuum level in $\approx 1.0$--7.0~keV ($N_\text{1.0--7.0~keV}$) vary with respect to the other spectral components (let this be $N_\text{1.0--7.0~keV}'$) depending on the count rate in 9.0--11.5~keV ($R_\text{9.0--11.5~keV}$) as
\begin{equation}
N_\text{1.0--7.0~keV}' = N_\text{1.0--7.0~keV} \, \left(\frac{R_\text{9.0--11.5~keV}}{R_{0}}\right)^{\alpha},
\end{equation}
where $R_{0}$ and $\alpha$ are free parameters.
Such a spectral variation is assumed to be due to the cosmic-ray spectral modulation in accordance with solar activity \citep[e.g.,][]{mizuno04, fiandrini20}, so that this is assumed to depend on a total particle-induced background rate.
The spectrum becomes harder in higher solar-activity periods, and this tendency is consistent with the cosmic-ray spectral modulation (e.g., \citealt{mizuno04, fiandrini20}).
Applying this model, we fit all the ACIS-stowed observation spectra simultaneously to determine the best-fit $R_{0}$ and $\alpha$ values for each CCD and each observation mode.
The resultant parameter values are presented in Table~\ref{tab-time}.
Applying this modification to the spectral models, we are able to describe better the observed hardness ratios (see Fig.~\ref{fig-hardness-vf} and \ref{fig-hardness-f}) and actual spectral shapes (see Fig.~\ref{fig-time2}), although this modeling is still not sufficient to fully explain the data.

\begin{table}[htb!]
\centering
\caption{The best-fit hardness-modification parameters $\alpha$ and $\beta$.$^{\rm a}$
\label{tab-time}}
\begin{threeparttable}
\begin{tabular}{l l l l}
\hline\hline
Data mode  &  CCD  &  $R_{0}$ (counts s$^{-1}$)  & $\alpha$  \\\hline
VFAINT   & S1 & 1.25 & 0.20 \\
  & S3 & 0.76 & 0.57 \\
  \hline
FAINT & S1 & 1.55 & 0.35 \\
  & S3 & 0.83 & 0.25 \\
\hline
\end{tabular}
\begin{tablenotes}
\item[a] As our purpose is to determine the best-fit parameters, we do not calculate their error ranges.
\end{tablenotes}
\end{threeparttable}
\end{table}

\section{Design of the Particle-Induced Background Spectral-Model Generation Tool ``{\tt mkacispback}''}\label{sec-tool}
Based on the spectral model configurations and their spatial and temporal variation parameters obtained in Section~\ref{sec-analysis}, a tool named ``{\tt mkacispback}'' has been developed to generate the particle-induced background spectral model for an arbitrary observation.
We use the 32 spectral models for each CCD and each data mode as ``template models'' for the {\tt mkacispback} tool.
In addition, the temporal variation parameters $R_{0}$ and $\alpha$ for each CCD and each observation mode are used.

\begin{figure*}[htb!]
\centering
\includegraphics[width=12cm]{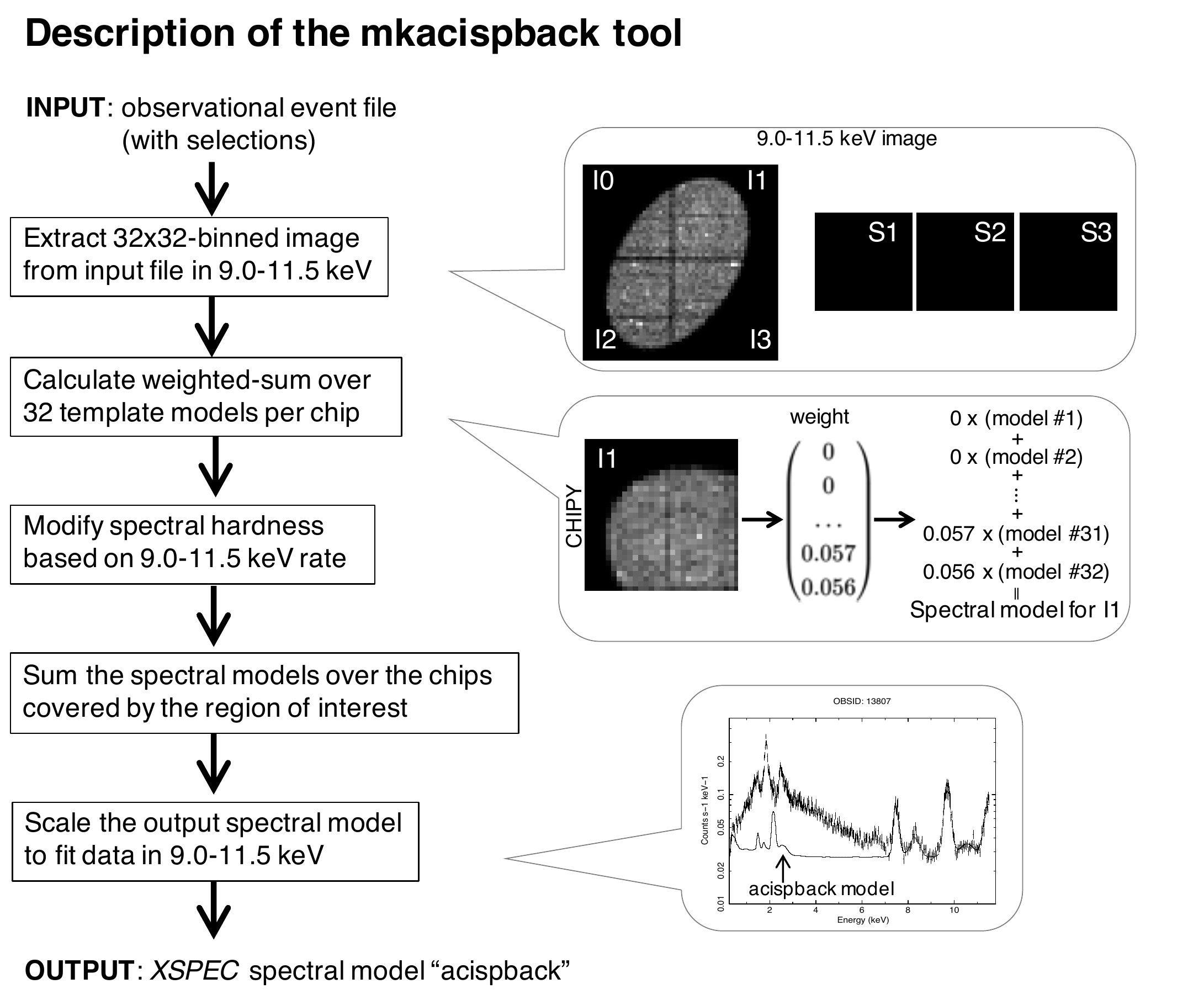}
\caption{Description of the process to generate the particle-induced background spectral model for an arbitrary observation data, which is taken in the {\tt mkacispback} tool.
\label{fig-toolflow}}
\end{figure*}

A brief description of the steps executed by the tool is presented in Fig.~\ref{fig-toolflow}.
The input data are the (level 2) events file and the spectral extraction regions.
First, in order to extract a particle-background count rate from the input region, it makes an image of the 9.0--11.5~keV energy band.
Second, it makes a ``weight map'' by dividing the image into $32\times 32$ regions per CCD.
This weight map is converted to a vector which contains 32 values corresponding to the 32 template models by taking sums over CHIPX.
Third, the total spectral model for one CCD is generated by taking a weighted-sum of the 32 template models based on this vector.
After getting the spectral model for each CCD, as the fourth step, it modifies the spectral hardness based on the count rate in the 9.0--11.5~keV energy range.
Fifth, the resultant spectral models for the CCDs covered by the region of interest are added together to obtain one spectral model.
Finally, this spectral model is scaled to fit the data in the 9.0--11.5~keV energy range to produce the final output.
This software is available at \url{https://github.com/hiromasasuzuki/mkacispback}.
The software is composed of shell scripts, Python (with {\it astropy} library), C++, CIAO tools, and FTOOLS.

\section{Verification of the {\tt mkacispback} Tool}\label{sec-verification}

\subsection{Comparison to the ACIS-stowed and blank-sky observations}\label{sec-verify}
In order to check the validity of our tool, we first apply this tool to the gain-corrected-and-merged ACIS-stowed and CDF-S blank-sky data.
For the S1 and S3 CCDs, because no data are available in the CDF-S blank-sky data set generated from the {\it Chandra} Deep Field South observations (Table~\ref{tab-blanksky}), we make use of the blank-sky data sets from CXC's CALDB\footnote{\url{https://cxc.cfa.harvard.edu/ciao/threads/acisbackground/}}.
We combine the CALDB blank-sky observations with the conditions, focal plane temperature lower than $-119^\circ$C, with CTI correction, and with TGAIN correction, and extract the spectra in VFAINT and FAINT modes.
The CALDB blank-sky data used in this work are summarized in Table~\ref{tab-caldb-blanksky}.

\begin{figure*}[htb!]
\centering
\includegraphics[width=16cm, ]{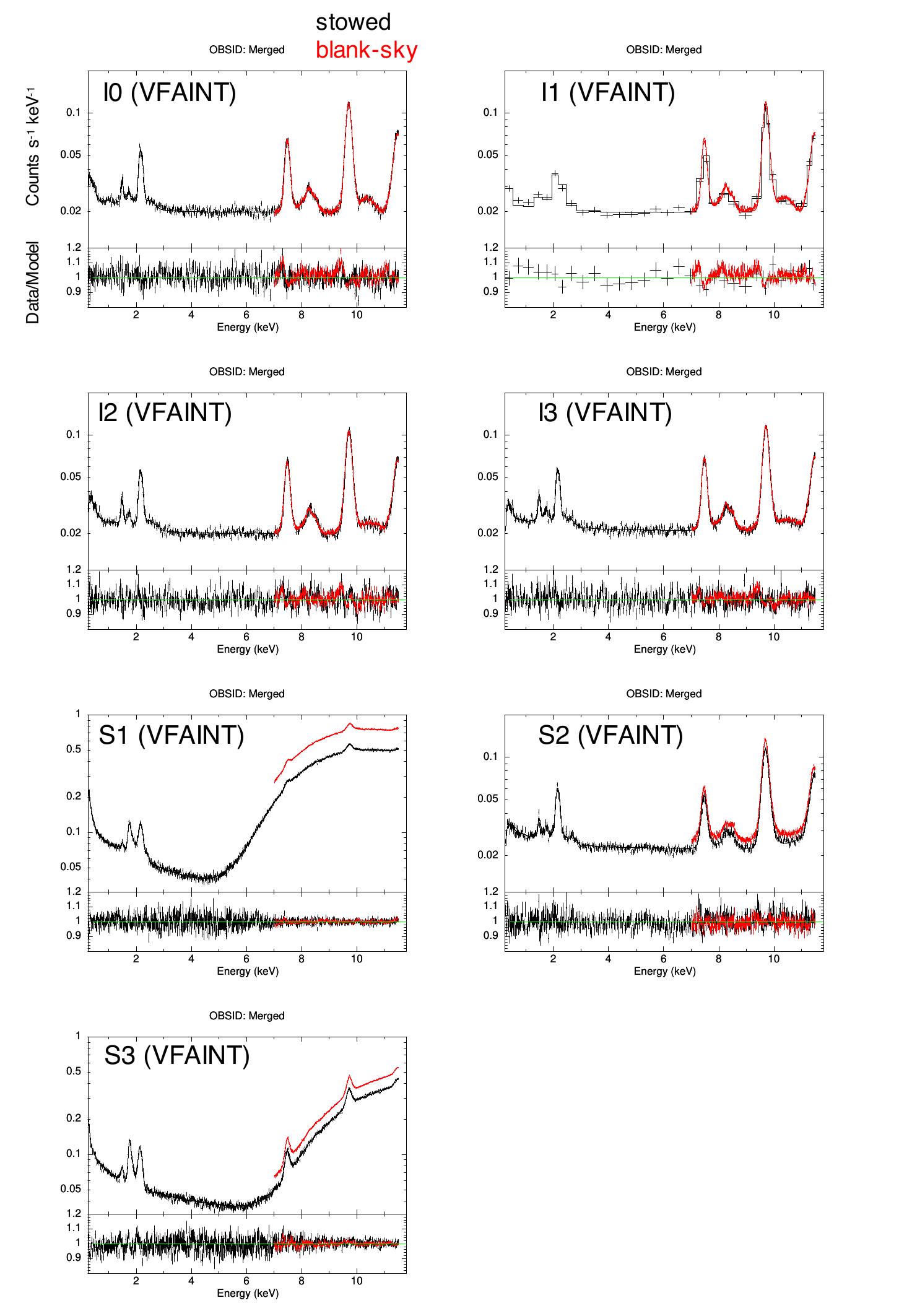}
\caption{Merged observational spectra obtained in VFAINT mode and the output spectral models of {\tt mkacispback}.
The spectra are extracted from the entire CCD regions.
The S1 and S3 data are taken from the CALDB blank-sky data instead of the CDF-S blank-sky data set.
For each CCD, the upper and lower panels represent the count rate and data-to-model ratio, respectively.
We note that the background was higher in the period covered by the CALDB blank-sky observations so that offsets between the ACIS-stowed and CALDB blank-sky data are present.
\label{fig-vfaint}}
\end{figure*}

\begin{figure*}[htb!]
\centering
\includegraphics[width=16cm]{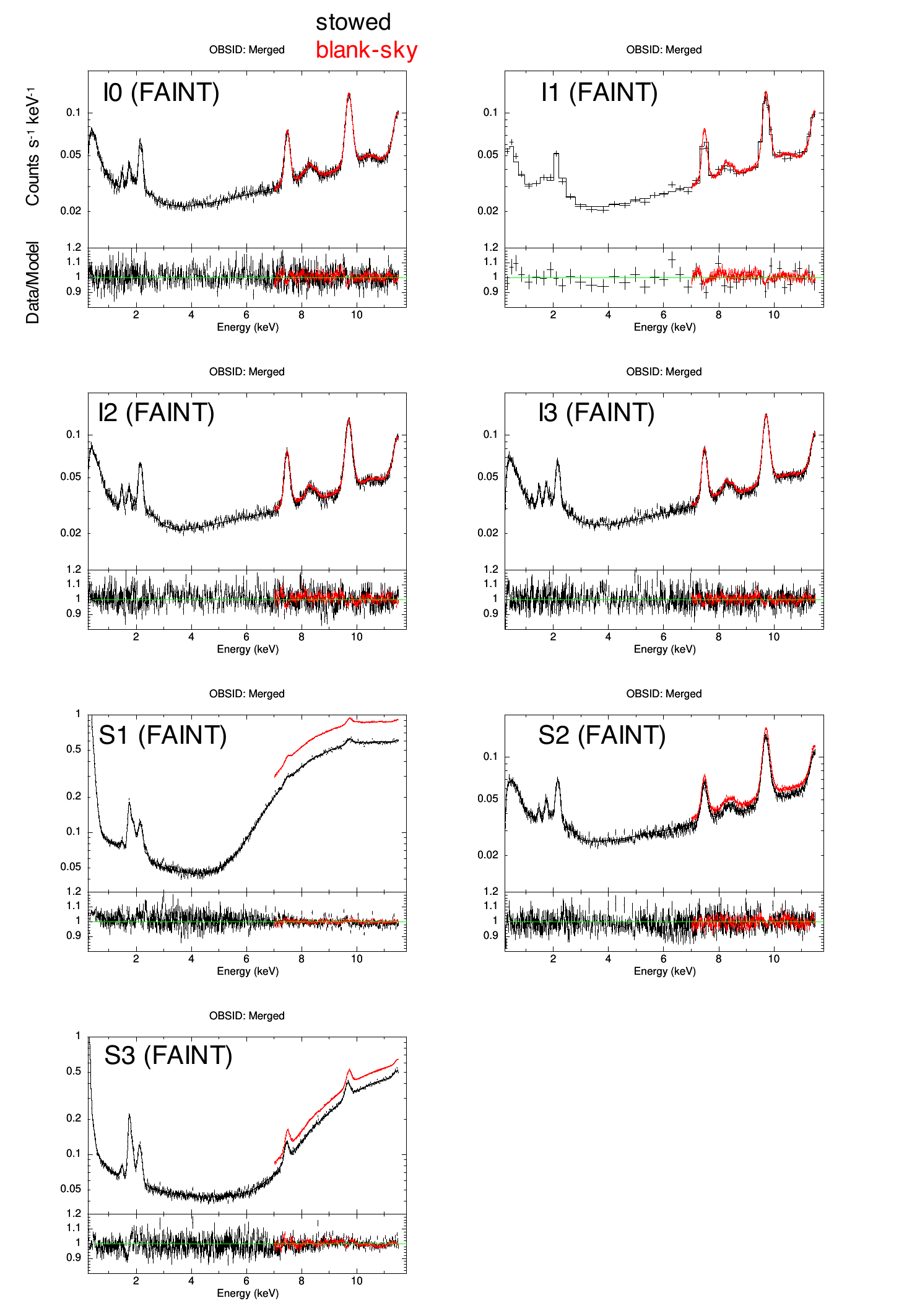}
\caption{Same as Fig.~\ref{fig-vfaint}, but for the data processed in FAINT mode.
\label{fig-faint}}
\end{figure*}

Figures~\ref{fig-vfaint} and \ref{fig-faint} present the comparisons of the data and output spectral models of {\tt mkacispback} for individual CCDs.
To examine the modeling accuracy of {\tt mkacispback}, we calculate the data-to-model ratio for the continuum regions (0.25--1.30~keV and 3.0--7.0~keV) and for the energies around strong lines (1.6--1.7~keV, 2.3--2.5~keV, 7.2--7.4~keV, and 9.2--9.6~keV).
The modeling accuracy for the continuum and line regions is found to be within 5\% and 8\%, respectively.
Judging from this, we conclude that the accuracy of the {\tt mkacispback} tool is sufficient for most applications.

We note several things about Figs.~\ref{fig-vfaint} and \ref{fig-faint}.
For the I1 CCD, although its spatial and temporal variations are parametrized with the same parameters as those for I0, the models represent the data pretty well.
At the energies below 0.7~keV, the S1 and S3 spectra in FAINT mode are higher than the models with discrepancies of $\lesssim 10\%$.
This will be either due to insufficient spectral modeling in Section~\ref{sec-analysis} or due to potential temporal/spatial variations which have not been treated in this work.
For several cases such as the S3, FAINT at $\approx 1.8$~keV and the I2, VFAINT around 9~keV, larger residuals of $\lesssim 10\%$ can be seen around the line components, which are probably due to insufficient gain corrections in Section~\ref{sec-analysis}.
These residuals can be addressed in a future update of {\tt mkacispback}.
It is worth noting that the differences in count rates seen in Figs.~\ref{fig-vfaint} and \ref{fig-faint} between the ACIS-stowed and CDF-S blank-sky data are due to different particle background levels during the different time intervals of the data sets. The model does a reasonable job of estimating this difference as seen in Figs.~\ref{fig-vfaint} and \ref{fig-faint}.

Figures~\ref{fig-vf-i0-eachobs}--\ref{fig-f-s3-eachobs} present the comparison between individual ACIS-stowed observations and output models of {\tt mkacispback}, to verify the model in more detail.
The spectra are extracted from the entire CCD regions.
As can be seen, for most cases, the models describe the data well without remarkable residual structures.
For some cases such as OBSIDs 62831, 62848, and 62850, relatively large residuals of $\lesssim 10\%$ even for their continuum regions may be present.
These may be due to our insufficient modeling of the temporal variations of the spectral shapes which can be inferred from Figs.~\ref{fig-hardness-vf} and \ref{fig-hardness-f} as well.
Future works may require detector simulations to understand the physical processes that are responsible for these variations.

\begin{table}[htb!]
\centering
\caption{CALDB blank-sky data list used in this work
\label{tab-caldb-blanksky}}
\begin{threeparttable}
\begin{tabular}{l l}
\hline\hline
CCD & Name \\ \hline
S1 & \url{acis5sD2000-12-01bkgrnd_ctiN0002.fits} \\
 & \url{acis5sD2005-09-01bkgrnd_ctiN0005.fits} \\
 & \url{acis5sD2009-09-21bkgrnd_ctiN0003.fits} \\
 & \url{acis5sD2012-01-01bkgrnd_ctiN0002.fits} \\
S3 & \url{acis7iD2000-12-01bkgrnd_ctiN0002.fits} \\
 & \url{acis7sD2000-12-01bkgrnd_ctiN0002.fits} \\
 & \url{acis7iD2005-09-01bkgrnd_ctiN0005.fits} \\
 & \url{acis7sD2005-09-01bkgrnd_ctiN0005.fits} \\
 & \url{acis7sD2009-09-21bkgrnd_ctiN0003.fits} \\
 & \url{acis7sD2012-01-01bkgrnd_ctiN0002.fits} \\
\hline
\end{tabular}
\begin{tablenotes}
\item[]
\end{tablenotes}
\end{threeparttable}
\end{table}

\subsection{Application to a celestial observation}
\begin{figure*}[htb!]
\centering
\includegraphics[width=16cm]{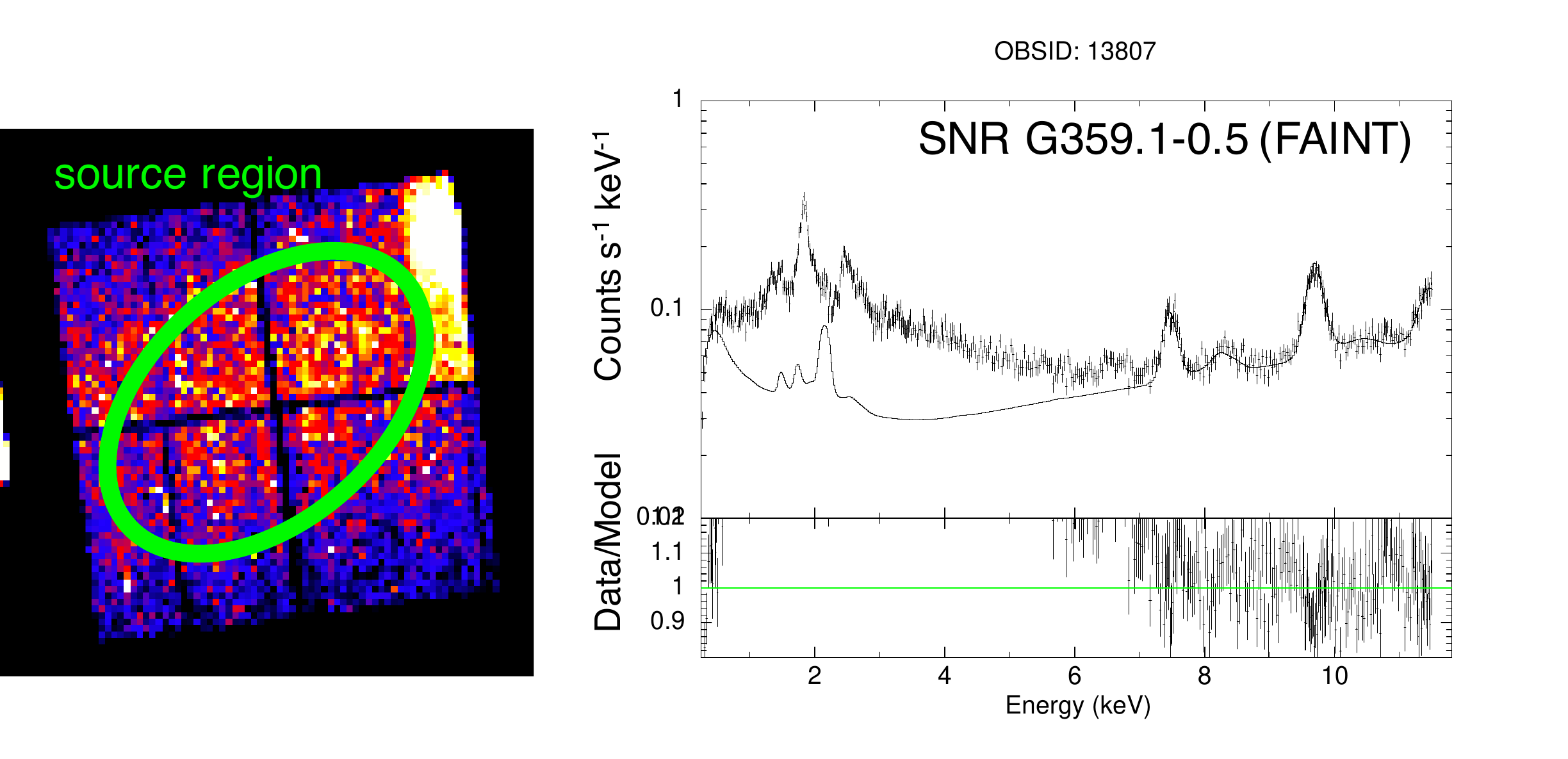}
\caption{An application of {\tt mkacispback} tool for the supernova remnant G359.1$-$0.5.
The left panel is the 0.7--5.0~keV image of the G359.1$-$0.5 region. The source region, which covers all the four ACIS-I CCDs, is shown with the green ellipse.
In the right panel, the extracted spectrum and the output spectral model of {\tt mkacispback}, as well as their residuals are shown.
The residuals below $\sim 7$~keV are large because a model for the source emission has not been included here.
\label{fig-g359}}
\end{figure*}

As a demonstration of the application of {\tt mkacispback} to actual celestial observations, we apply this tool to the X-ray emission of the supernova remnant G359.1-0.5, which is relatively old among supernova remnants and thus is a faint and diffuse source, where the particle-induced background is relatively important (e.g., \citealt{suzuki20a}).
An image of G359.1$-$0.5 is shown in the left panel of Fig.~\ref{fig-g359} with the spectral extraction region indicated by the green ellipse.
The source spectrum with the background spectrum produced by {\tt mkacispback} is shown in the right panel of Fig.~\ref{fig-g359}.
As well as the high-energy range of the 7.0--11.5~keV, the very low-energy part of $\lesssim 0.5$~keV also shows a reasonable match between the data and model.
We have shown an example application for an observation of a supernova remnant but the tool should work well for observations of other extended sources, such as clusters of galaxies.

\begin{figure*}[htb!]
\centering
\includegraphics[width=16cm]{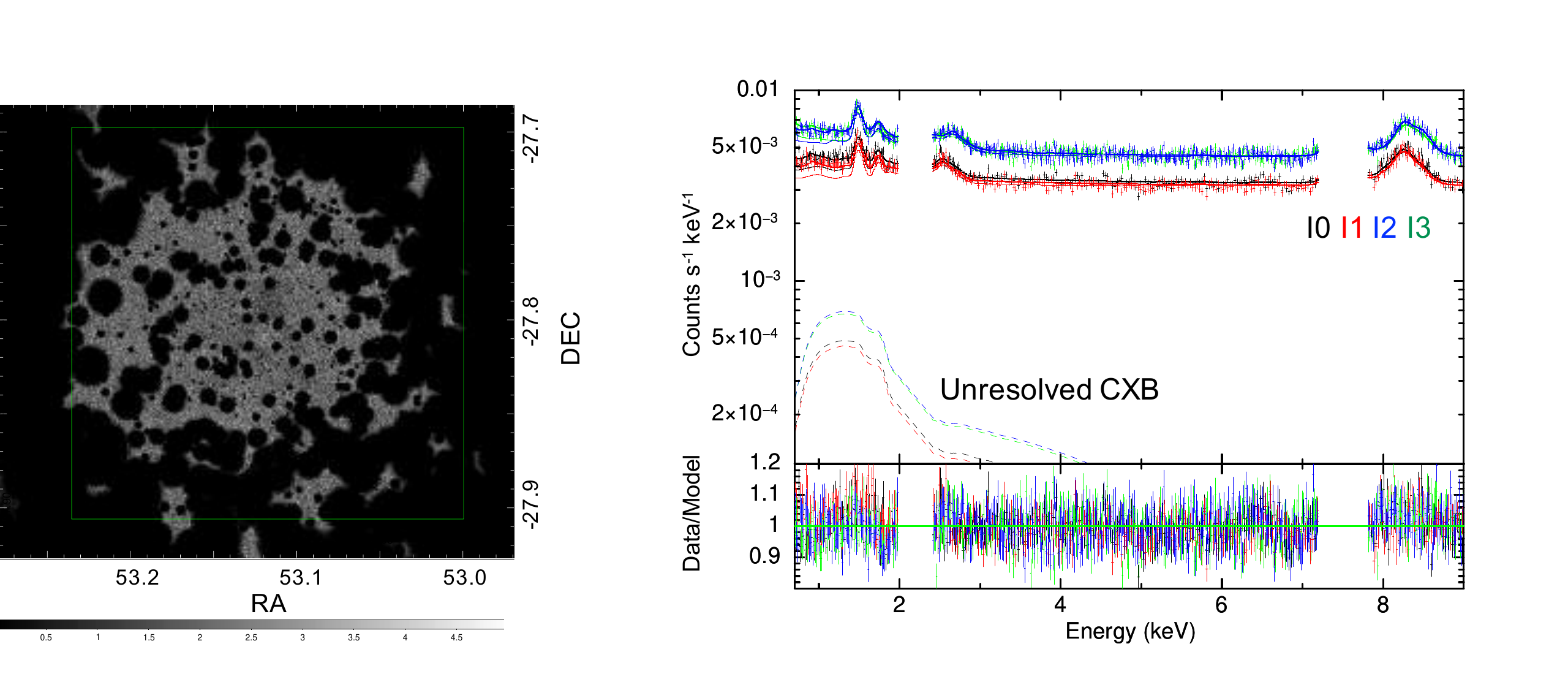}
\caption{Left panel shows the entire CDF-S image after point-source removal. Green square is the spectral extraction region.
Right panel represents the spectral fitting result. Data (crosses) and best-fit models (thick solid lines) are shown for four ACIS-I CCDs.
The upper and lower panels indicate the count rate and data-to-model ratio, respectively.
The particle-induced background models are shown with thin solid lines.
The unresolved CXB models are indicated with dashed lines.
The 0.7--9.0~keV range is used for the spectral fit. Two energy ranges, 2.0--2.4~keV and 7.2--7.8~keV, are excluded from analysis.
\label{fig-cxb}}
\end{figure*}

\subsection{Estimation of the unresolved intensity of the Cosmic X-ray Background}
As a validation and application of {\tt mkacispback}, here we evaluate the unresolved intensity of the Cosmic X-ray Background (CXB).
Overall, we follow the analysis method adopted in Sections~4 and 5 of \cite{bartalucci14} with an updated point source catalog in the CDF-S region by \cite{luo17}.
The data reduction is done as follows: cataloged point sources identified by \cite{luo17} are removed from individual CDF-S observations. The exclusion regions consist of circles with radii ($r$) depending on off-axis angle ($\theta$) and source fluxes ($f$: photon flux in the 0.5--8.0~keV energy range). The $r$ is defined as
\begin{equation}
r \,(\arcsec) = C_f \,\left[ 1 + 10 \left( \frac{\theta \,(\arcsec)}{600} \right) \right],
\end{equation}
where $C_f$ is a scaling factor. The $C_f$ is set to be 2, 4.5, 6, and 9 for $f$ of $f < 0.02 \times 10^{-3}$ cnt s$^{-1}$, $0.02 \times 10^{-3} < f < 0.2 \times 10^{-3}$ cnt s$^{-1}$, $0.2 \times 10^{-3} < f < 2 \times 10^{-3}$ cnt s$^{-1}$, and $f > 2 \times 10^{-3}$ cnt s$^{-1}$, respectively.
This exclusion is applied to the individual CDF-S observations listed in Table~\ref{tab-blanksky}.
The analysis region is defined as a $12\farcm5 \times 12\farcm5$ square centered on R.A.(2000) $= 53\fdg1175$, decl.(2000) $= -27\fdg8019$.
The spectrum is extracted from the analysis region for the entire CDF-S data set after the point-source exclusion.

To get the particle-induced background model suited for the analysis region, we run {\tt mkacispback} for each CDF-S observation and take an exposure-and-area-weighted sum of the model spectra.
Following \cite{bartalucci14}, the total spectral model to be compared to the observationa is assumed to be 
\begin{equation}
{\rm Abs.} \times ({\rm LP} + {\rm HP} + {\rm CXB_{UR}}) + (\text{particle-induced background}),
\end{equation}
where Abs. is the Galactic absorption fixed to $8.8 \times 10^{19}$ cm$^{-2}$ \citep{stark92} modeled with {\tt tbabs} in XSPEC,
and LP, HP, and CXB$_{\rm UR}$ represent lower-temperature (0.14~keV), higher-temperature (0.248~keV) thermal plasmas modeled with {\tt apec} in XSPEC with solar abundances, and the unresolved CXB modeled with {\tt powerlaw} in XSPEC, respectively.
The temperatures of the LP and HP models are fixed to the values obtained in \cite{bartalucci14}, and only their normalizations are treated as free parameters.
For the CXB$_{\rm UR}$ model, the spectral index of 1.42 and a free normalization are assumed.

The analysis region is shown on the image after point sources have been removed in the left panel of Fig.~\ref{fig-cxb}.
We extract spectra from the individual ACIS-I CCDs, and perform a simultaneous spectral fit for them.
In the spectral fit, we exclude two energy ranges near the line emission around 2.2 and 7.5~keV as there are significant residuals in these regions (see Section~\ref{sec-verify}).
Excluding these regions results in a better constraint on the normalization of the CXB$_{\rm UR}$ component, which is the parameter of interest in this fit.
The spectral fit is shown in the right panel of Fig.~\ref{fig-cxb}.
The estimated unresolved CXB intensities (in erg s$^{-1}$ cm$^{-2}$ deg$^{-2}$) are 3.10 (2.98--$3.21)\times 10^{-12}$ in the 2--8~keV band, and 8.35 (8.00--$8.70) \times 10^{-12}$ in the 1--2~keV band.
These values are consistent with or lower than \cite{hickox06}, \cite{bartalucci14}, and \cite{luo17}.\footnote{For example, our estimates are lower than those of \cite{bartalucci14} by 10--20\%. Such differences are probably due to the updated point-source catalog and slight difference in particle-induced background models.}

\section{Conclusion}
In this work, we investigated the particle-induced background properties of the {\it Chandra} ACIS-I and S1, S2, and S3 CCDs, and for both of the two data modes, FAINT and VFAINT.
We used the ACIS-stowed and CDF-S blank-sky data sets to obtain spectral models to describe the background.
We found and modeled the temporal variation of the spectral normalizations and shapes for the first time, as well as the spatial variations along the CHIPY axis by dividing each CCD into 32 regions in the CHIPY direction.
The spectral hardness was found to vary in accordance with the total flux for the BI CCDs.
Combining these temporal and spatial parameterizations, we have developed a tool ``mkacispback'' to generate the particle-induced background spectrum for an arbitrary observation.
This tool was verified using the ACIS-stowed and CDF-S/CALDB blank-sky observations and was found to be reliable within 5\% in the continuum and 8\% around the lines.
As a verification and application of our models, we also evaluated the unresolved CXB intensities as 3.10 (2.98--$3.21)\times 10^{-12}$ erg s$^{-1}$ cm$^{-2}$ deg$^{-2}$ in the 2--8~keV band, and 8.35 (8.00--$8.70) \times 10^{-12}$ erg s$^{-1}$ cm$^{-2}$ deg$^{-2}$ in the 1--2~keV band using {\tt mkacispback} and the CDF-S observations. These estimates are consistent with or lower than previous ones.
This tool is available at \url{https://github.com/hiromasasuzuki/mkacispback}.

\begin{acknowledgements}

We are grateful to the $Chandra$ operation team and the IACHEC team for their patient operation, maintenance, and calibration of $Chandra$.
We acknowledge the help by Catharine Grant in understanding the CHIPY-$\Delta E$ plots.
H.S. appreciate the supports of the people at the Center for Astrophysics | Harvard-Smithsonian during my stay, which enabled this work.
H.S. is supported by JSPS Research Fellowship for Young Scientists (Nos.~19J11069 and 21J00031) and Overseas Challenge Program for Young Researchers (No. 201980289).
T.J.G. and P.P.P. acknowledge support under NASA contract NAS8-03060 with the {\it Chandra} X-ray Center.

\end{acknowledgements}

\bibliography{acispback}

\begin{thebibliography}{23}
\expandafter\ifx\csname natexlab\endcsname\relax\def\natexlab#1{#1}\fi

\bibitem[{{Arnaud}(1996)}]{arnaud96}
{Arnaud}, K.~A. 1996, in Astronomical Society of the Pacific Conference Series,
  Vol. 101, Astronomical Data Analysis Software and Systems V, ed. G.~H.
  {Jacoby} \& J.~{Barnes}, 17

\bibitem[{{Bartalucci} {et~al.}(2014){Bartalucci}, {Mazzotta}, {Bourdin}, \&
  {Vikhlinin}}]{bartalucci14}
{Bartalucci}, I., {Mazzotta}, P., {Bourdin}, H., \& {Vikhlinin}, A. 2014, \aap,
  566, A25

\bibitem[{Bearden {et~al.}(1967)Bearden, Burr, \& States.}]{bearden67}
Bearden, J.~A., Burr, A.~F., \& States., U. 1967, X-ray wavelengths and x-ray
  atomic energy levels [electronic resource] / J.A. Bearden (U.S. Dept. of
  Commerce, National Bureau of Standards : For sale by the Supt. of Docs., U.S.
  G.P.O Washington, D.C), 66 p. ;

\bibitem[{{Cash}(1979)}]{cash79}
{Cash}, W. 1979, \apj, 228, 939

\bibitem[{{Fiandrini} {et~al.}(2020){Fiandrini}, {Tomassetti}, {Bertucci},
  {Donnini}, {Graziani}, \& {Khiali}}]{fiandrini20}
{Fiandrini}, E., {Tomassetti}, N., {Bertucci}, B., {et~al.} 2020, arXiv
  e-prints, arXiv:2010.08649

\bibitem[{Fruscione {et~al.}(2006)Fruscione, McDowell, Allen, Brickhouse,
  Burke, Davis, Durham, Elvis, Galle, Harris, Huenemoerder, Houck, Ishibashi,
  Karovska, Nicastro, Noble, Nowak, Primini, Siemiginowska, Smith, \&
  Wise}]{fruscione06}
Fruscione, A., McDowell, J.~C., Allen, G.~E., {et~al.} 2006, in Observatory
  Operations: Strategies, Processes, and Systems, ed. D.~R. Silva \& R.~E.
  Doxsey, Vol. 6270, International Society for Optics and Photonics (SPIE), 586
  -- 597

\bibitem[{{Gastaldello} {et~al.}(2017){Gastaldello}, {Ghizzardi}, {Marelli},
  {Salvetti}, {Molendi}, {De Luca}, {Moretti}, {Rossetti}, \&
  {Tiengo}}]{gastaldello17}
{Gastaldello}, F., {Ghizzardi}, S., {Marelli}, M., {et~al.} 2017, Experimental
  Astronomy, 44, 321

\bibitem[{{Grant} {et~al.}(2020){Grant}, {Miller}, {Bautz}, {Eraerds},
  {Molendi}, {Keelan}, {Hall}, {Holland}, {Kraft}, {Bulbul}, {Nulsen}, \&
  {Allen}}]{grant20}
{Grant}, C.~E., {Miller}, E.~D., {Bautz}, M.~W., {et~al.} 2020, in Society of
  Photo-Optical Instrumentation Engineers (SPIE) Conference Series, Vol. 11444,
  Society of Photo-Optical Instrumentation Engineers (SPIE) Conference Series,
  1144442

\bibitem[{{Hagino} {et~al.}(2020){Hagino}, {Odaka}, {Sato}, {Sato}, {Suzuki},
  {Mizuno}, {Kawaharada}, {Ohno}, {Nakazawa}, {Kobayashi}, {Murakami},
  {Miyake}, {Asai}, {Koi}, {Madejski}, {Saito}, {Wright}, {Enoto}, {Fukazawa},
  {Hayashi}, {Kataoka}, {Katsuta}, {Kokubun}, {Laurent}, {Lebrun}, {Limousin},
  {Maier}, {Makishima}, {Mori}, {Nakamori}, {Nakano}, {Noda}, {Ohta}, {Sato},
  {Tajima}, {Takahashi}, {Takahashi}, {Takeda}, {Tanaka}, {Terada}, {Uchiyama},
  {Uchiyama}, {Watanabe}, {Yamaoka}, {Yatsu}, \& {Yuasa}}]{hagino20}
{Hagino}, K., {Odaka}, H., {Sato}, G., {et~al.} 2020, Journal of Astronomical
  Telescopes, Instruments, and Systems, 6, 046003

\bibitem[{{Hickox} \& {Markevitch}(2006)}]{hickox06}
{Hickox}, R.~C. \& {Markevitch}, M. 2006, \apj, 645, 95

\bibitem[{{Kaastra}(2017)}]{kaastra17}
{Kaastra}, J.~S. 2017, \aap, 605, A51

\bibitem[{{Kuntz} \& {Snowden}(2008)}]{kuntz08}
{Kuntz}, K.~D. \& {Snowden}, S.~L. 2008, \aap, 478, 575

\bibitem[{{Lotti} {et~al.}(2017){Lotti}, {Mineo}, {Jacquey}, {Molendi},
  {D'Andrea}, {Macculi}, \& {Piro}}]{lotti17}
{Lotti}, S., {Mineo}, T., {Jacquey}, C., {et~al.} 2017, Experimental Astronomy,
  44, 371

\bibitem[{{Luo} {et~al.}(2017){Luo}, {Brandt}, {Xue}, {Lehmer}, {Alexander},
  {Bauer}, {Vito}, {Yang}, {Basu-Zych}, {Comastri}, {Gilli}, {Gu},
  {Hornschemeier}, {Koekemoer}, {Liu}, {Mainieri}, {Paolillo}, {Ranalli},
  {Rosati}, {Schneider}, {Shemmer}, {Smail}, {Sun}, {Tozzi}, {Vignali}, \&
  {Wang}}]{luo17}
{Luo}, B., {Brandt}, W.~N., {Xue}, Y.~Q., {et~al.} 2017, \apjs, 228, 2

\bibitem[{{Marelli} {et~al.}(2021){Marelli}, {Molendi}, {Rossetti},
  {Gastaldello}, {Salvetti}, {De Luca}, {Bartalucci}, {K{\"u}hl}, {Esposito},
  {Ghizzardi}, \& {Tiengo}}]{marelli21}
{Marelli}, M., {Molendi}, S., {Rossetti}, M., {et~al.} 2021, \apj, 908, 37

\bibitem[{{Markevitch} {et~al.}(2003){Markevitch}, {Bautz}, {Biller}, {Butt},
  {Edgar}, {Gaetz}, {Garmire}, {Grant}, {Green}, {Juda}, {Plucinsky},
  {Schwartz}, {Smith}, {Vikhlinin}, {Virani}, {Wargelin}, \&
  {Wolk}}]{markevitch03}
{Markevitch}, M., {Bautz}, M.~W., {Biller}, B., {et~al.} 2003, \apj, 583, 70

\bibitem[{{Mizuno} {et~al.}(2004){Mizuno}, {Kamae}, {Godfrey}, {Handa},
  {Thompson}, {Lauben}, {Fukazawa}, \& {Ozaki}}]{mizuno04}
{Mizuno}, T., {Kamae}, T., {Godfrey}, G., {et~al.} 2004, \apj, 614, 1113

\bibitem[{{NASA High Energy Astrophysics Science Archive Research Center
  (HEASARC)}(2014)}]{heasarc14}
{NASA High Energy Astrophysics Science Archive Research Center (HEASARC)}.
  2014, {HEAsoft: Unified Release of FTOOLS and XANADU}

\bibitem[{{Salvetti} {et~al.}(2017){Salvetti}, {Marelli}, {Gastaldello},
  {Ghizzardi}, {Molendi}, {De Luca}, {Moretti}, {Rossetti}, \&
  {Tiengo}}]{salvetti17}
{Salvetti}, D., {Marelli}, M., {Gastaldello}, F., {et~al.} 2017, Experimental
  Astronomy, 44, 309

\bibitem[{{Stark} {et~al.}(1992){Stark}, {Gammie}, {Wilson}, {Bally}, {Linke},
  {Heiles}, \& {Hurwitz}}]{stark92}
{Stark}, A.~A., {Gammie}, C.~F., {Wilson}, R.~W., {et~al.} 1992, \apjs, 79, 77

\bibitem[{{Suzuki} {et~al.}(2020){Suzuki}, {Bamba}, {Enokiya}, {Yamaguchi},
  {Plucinsky}, \& {Odaka}}]{suzuki20a}
{Suzuki}, H., {Bamba}, A., {Enokiya}, R., {et~al.} 2020, \apj, 893, 147

\bibitem[{{Tawa} {et~al.}(2008){Tawa}, {Hayashida}, {Nagai}, {Nakamoto},
  {Tsunemi}, {Yamaguchi}, {Ishisaki}, {Miller}, {Mizuno}, {Dotani}, {Ozaki}, \&
  {Katayama}}]{tawa08}
{Tawa}, N., {Hayashida}, K., {Nagai}, M., {et~al.} 2008, \pasj, 60, S11

\bibitem[{{Wik} {et~al.}(2014){Wik}, {Hornstrup}, {Molendi}, {Madejski},
  {Harrison}, {Zoglauer}, {Grefenstette}, {Gastaldello}, {Madsen},
  {Westergaard}, {Ferreira}, {Kitaguchi}, {Pedersen}, {Boggs}, {Christensen},
  {Craig}, {Hailey}, {Stern}, \& {Zhang}}]{wik14}
{Wik}, D.~R., {Hornstrup}, A., {Molendi}, S., {et~al.} 2014, \apj, 792, 48

\end{thebibliography}

\appendix
\onecolumn
\section{Logs of the ACIS-stowed and CDF-S blank-sky observations}

\small
\begin{ThreePartTable}
\begin{TableNotes}
\item[a] All observations use ACIS-I0, I2, I3, and S1--S3 except OBSID 62678, which uses ACIS-I0--I3, S1 and S3.
\end{TableNotes}

\begin{longtable}[htb!]{l l l}
\caption{ACIS-stowed observation logs\tnote{a}.\label{tab-stowed}}\\
\hline\hline
Observation ID & Date & Exposure (ksec)  \\
\hline
\endfirsthead
\hline\hline
Observation ID & Date & Exposure (ksec)  \\
\hline
\endhead
\endfoot
\insertTableNotes
\endlastfoot

62850  &  2002-09-03  &  52.49 \\
62848  &  2003-05-04  &  47.46 \\
62846  &  2003-12-08  &  45.86 \\
62836  &  2004-11-04  &  46.62 \\
62831  &  2005-06-10  &  47.20 \\
62824  &  2005-11-13  &  47.17 \\
62823  &  2006-06-01  &  44.11 \\
62819  &  2006-11-18  &  47.29 \\
62816  &  2007-05-28  &  46.28 \\
62815  &  2007-11-08  &  46.45 \\
62814  &  2008-07-26  &  48.72 \\
62813  &  2008-11-03  &  49.02 \\
62812  &  2009-06-18  &  33.15 \\
62811  &  2009-06-19  &  12.38 \\
62810  &  2009-11-04  &  49.86 \\
62809  &  2010-03-06  &  45.60 \\
62808  &  2010-11-12  &  47.38 \\
62804  &  2011-07-12  &  47.45 \\
62802  &  2011-11-09  &  50.56 \\
62678  &  2012-06-15  &  47.38  \\
62668  &  2015-12-09  &  47.38 \\
62667  &  2016-09-26  &  10.64 \\
62666  &  2016-10-14  &  9.00 \\
62665  &  2016-11-03  &  9.10 \\
62664  &  2016-12-08  &  9.11 \\
\hline

\end{longtable}
\end{ThreePartTable}
\normalsize

\small
\begin{ThreePartTable}
\begin{TableNotes}
\item[a] A very early observation (OBSID: 1431, taken on 1999-10-15) is excluded because of the high focal plane temperature ($-109^\circ$C).
The observations in which the operations were done in FAINT mode (OBSIDs: 441, 582, 1672, 2239, 2312, 2313, 2405, 2406, 2409) are also excluded.
One observation (OBSID: 17542) exhibits a flare-like behavior and is excluded.
\item[b] Data for S1 and S3 are not available in the CDF-S blank-sky observations.
\end{TableNotes}

\begin{longtable}[htb!]{l l l c}
\caption{CDF-S blank-sky observation logs.\tnote{a}\label{tab-blanksky}}\\
\hline\hline
Observation ID & Date & Exposure (ksec)  & Note\tnote{b} \\
\hline
\endfirsthead
\hline\hline
Observation ID & Date & Exposure (ksec)  & Note\tnote{b} \\
\hline
\endhead
\endfoot
\insertTableNotes
\endlastfoot

8591  &  2007-09-20  &  45.43 \\
9593  &  2007-09-22  &  46.43 \\
9718  &  2007-10-03  &  49.38 \\
8593  &  2007-10-06  &  49.49 \\
8597  &  2007-10-17  &  59.28 \\
8595  &  2007-10-19  &  115.42 \\
8592  &  2007-10-22  &  86.64 \\
8596  &  2007-10-24  &  115.12 \\
9575  &  2007-10-27  &  108.69 \\
9578  &  2007-10-30  &  38.57 \\
8594  &  2007-11-01  &  141.40 \\
9596  &  2007-11-04  &  111.89 \\
12043  &  2010-03-18  &  129.58 \\
12123  &  2010-03-21  &  24.79 \\
12044  &  2010-03-23  &  99.53 \\
12128  &  2010-03-27  &  22.80 \\
12045  &  2010-03-28  &  99.72 \\
12129  &  2010-04-03  &  77.14 \\
12135  &  2010-04-06  &  62.53 \\
12046  &  2010-04-08  &  78.02 \\
12047  &  2010-04-12  &  10.14 \\
12137  &  2010-04-16  &  92.78 \\
12138  &  2010-04-18  &  38.53 \\
12055  &  2010-05-15  &  80.68 \\
12213  &  2010-05-17  &  61.29 \\
12048  &  2010-05-23  &  138.10 \\
12049  &  2010-05-28  &  86.94 \\
12050  &  2010-06-03  &  29.66 \\
12222  &  2010-06-05  &  30.64 \\
12219  &  2010-06-06  &  33.66 \\
12051  &  2010-06-10  &  57.29 \\
12218  &  2010-06-11  &  87.98 \\
12223  &  2010-06-13  &  100.71 \\
12052  &  2010-06-15  &  110.41 \\
12220  &  2010-06-18  &  48.13 \\
12053  &  2010-07-05  &  68.11 \\
12054  &  2010-07-09  &  61.00 \\
12230  &  2010-07-11  &  33.81 \\
12231  &  2010-07-12  &  24.72 \\
12227  &  2010-07-14  &  54.32 \\
12233  &  2010-07-16  &  35.57 \\
12232  &  2010-07-18  &  32.89 \\
12234  &  2010-07-22  &  49.15 \\
16183  &  2014-06-09  &  98.78  & No S2 data \\
16180  &  2014-06-22  &  49.44  & No S2 data \\
16456  &  2014-07-29  &  47.46  & No S2 data \\
16641  &  2014-07-31  &  46.53  & No S2 data \\
16457  &  2014-08-05  &  45.98  & No S2 data \\
16644  &  2014-08-06  &  44.01  & No S2 data \\
16463  &  2014-09-23  &  53.22  & No S2 data \\
17417  &  2014-09-25  &  12.67  & No S2 data \\
17416  &  2014-09-28  &  52.40  & No S2 data \\
16454  &  2014-10-01  &  47.13  & No S2 data \\
16176  &  2014-10-02  &  24.68  & No S2 data \\
16175  &  2014-10-03  &  53.09  & No S2 data \\
16178  &  2014-10-07  &  73.98  & No S2 data \\
16177  &  2014-10-08  &  126.41  & No S2 data \\
16620  &  2014-10-10  &  33.71  & No S2 data \\
16462  &  2014-10-14  &  143.91  & No S2 data \\
17535  &  2014-10-17  &  121.92  & No S2 data \\
16184  &  2014-10-26  &  55.38  & No S2 data \\
16182  &  2014-10-28  &  75.06  & No S2 data \\
16181  &  2014-10-31  &  70.28  & No S2 data \\
17546  &  2014-11-02  &  19.82  & No S2 data \\
16186  &  2014-11-02  &  29.40  & No S2 data \\
16187  &  2014-11-03  &  28.07  & No S2 data \\
16188  &  2014-11-13  &  103.70  & No S2 data \\
16450  &  2014-11-18  &  81.04  & No S2 data \\
16190  &  2014-11-22  &  116.74  & No S2 data \\
16189  &  2014-11-29  &  90.46  & No S2 data \\
17556  &  2014-12-09  &  46.87  & No S2 data \\
16179  &  2014-12-31  &  29.98  & No S2 data \\
17573  &  2015-01-04  &  39.57  & No S2 data \\
17633  &  2015-03-16  &  35.52  & No S2 data \\
17634  &  2015-03-19  &  9.27  & No S2 data \\
16453  &  2015-03-21  &  70.28  & No S2 data \\
16451  &  2015-03-24  &  112.60  & No S2 data \\
16461  &  2015-05-19  &  129.37  & No S2 data \\
16191  &  2015-05-25  &  83.74  & No S2 data \\
16460  &  2015-06-16  &  21.44  & No S2 data \\
16459  &  2015-06-20  &  72.12  & No S2 data \\
17552  &  2015-10-10  &  49.91  & No S2 data \\
16455  &  2015-10-27  &  89.60  & No S2 data \\
16458  &  2015-10-30  &  96.38  & No S2 data \\
17677  &  2015-11-15  &  108.73  & No S2 data \\
18709  &  2015-11-22  &  16.86  & No S2 data \\
18719  &  2015-12-10  &  34.53  & No S2 data \\
16452  &  2015-12-12  &  27.71  & No S2 data \\
18730  &  2016-02-02  &  29.68  & No S2 data \\
16185  &  2016-03-24  &  48.43  & No S2 data \\
\hline

\end{longtable}
\end{ThreePartTable}
\normalsize

\section{Example of the base spectral models}

An explicit spectral model for CHIPY=993:1024 of I0 in VFAINT mode is presented here as an example.
The entire XSPEC model is described as \\
{\tt gaussian + gaussian + gaussian + gaussian + gaussian + gaussian + gaussian + gaussian + gaussian + gaussian + gaussian + gaussian + gaussian + gaussian + gaussian + fsline + fsline + fsline + fsline + gabs*powerlaw + gabs*constant*expdec + gaussian}.\\
The {\tt fsline} is a user-defined model for the frame-store lines, which is described by Eq.~\ref{eq-fsline}.
The model parameters are summarized in Table~\ref{tab-model}.
See Table~\ref{tab-lines} for the line identifications and gaussian line energies in Table~\ref{tab-model}.
Note that the models to describe continuum are purely phenomenological and most of the parameters are physically meaningless.
All the base models including all the CCDs and data modes can be found on \url{https://github.com/hiromasasuzuki/mkacispback}.

\begin{table}[htb!]
\fontsize{7}{9}\selectfont
\centering
\caption{Parameters of the analytical spectral model for CHIPY=993:1024, I0, VFAINT mode
\label{tab-model}}
\begin{threeparttable}
\begin{tabular}{l l l }
\hline\hline
{\tt gaussian} & Energy (keV) & 1.48656 \\
   & Sigma (keV) & 0 \\
   & Norm & 2.78059e-20 \\
{\tt gaussian} & Energy (keV) & 1.55745 \\
   & Sigma (keV) & 0 \\
   & Norm & 6.53328e-23 \\
{\tt gaussian} & Energy (keV) & 1.73978 \\
   & Sigma (keV) & 0 \\
   & Norm & 1.12338e-20 \\
{\tt gaussian} & Energy (keV) & 2.118 \\
   & Sigma (keV) & 0 \\
   & Norm & 2.65067e-06 \\
{\tt gaussian} & Energy (keV) & 2.1229 \\
   & Sigma (keV) & 0 \\
   & Norm & 0.00320472 \\
{\tt gaussian} & Energy (keV) & 2.205 \\
   & Sigma (keV) & 0 \\
   & Norm & 0.00422509 \\
{\tt gaussian} & Energy (keV) & 2.41 \\
   & Sigma (keV) & 0 \\
   & Norm & 2.11573e-21 \\
{\tt gaussian} & Energy (keV) & 7.4609 \\
   & Sigma (keV) & 0 \\
   & Norm & 0.00904457 \\
{\tt gaussian} & Energy (keV) & 7.4782 \\
   & Sigma (keV) & 0 \\
   & Norm & 1.92779e-19 \\
{\tt gaussian} & Energy (keV) & 8.2647 \\
   & Sigma (keV) & 0 \\
   & Norm & 0.00406398 \\
{\tt gaussian} & Energy (keV) & 8.4939 \\
   & Sigma (keV) & 0 \\
   & Norm & 0.0017254 \\
{\tt gaussian} & Energy (keV) & 9.7133 \\
   & Sigma (keV) & 0 \\
   & Norm & 0.0133757 \\
{\tt gaussian} & Energy (keV) & 9.628 \\
   & Sigma (keV) & 0 \\
   & Norm & 0.011449 \\
{\tt gaussian} & Energy (keV) & 11.5847 \\
   & Sigma (keV) & 0 \\
   & Norm & 0.00468481 \\
{\tt gaussian} & Energy (keV) & 11.4423 \\
   & Sigma (keV) & 0 \\
   & Norm & 0.00908103 \\
{\tt fsline} & $E_{\rm min}$ (keV) & 1.52664  \\
   & $E_{\rm max}$ (keV) & 1.53534 \\
   & Norm & 0.180872 \\
{\tt fsline} & $E_{\rm min}$ (keV) & 2.50853 \\
   & $E_{\rm max}$ (keV) & 2.80625  \\
   & Norm & 0.0109215 \\
{\tt fsline} & $E_{\rm min}$ (keV) & 8.04112 \\
   & $E_{\rm max}$ (keV) & 8.04124 \\
   & Norm & 1.5326e+07 \\
{\tt fsline} & $E_{\rm min}$ (keV) & 10.5649 \\
   & $E_{\rm max}$ (keV) & 10.8799 \\
   & Norm & 0.0196257 \\
{\tt gabs} & Energy (keV) & 0.158332 \\
   & Sigma (keV) & 0.024892 \\
   & Norm & 999963 \\
{\tt powerlaw} & Index & 0.0221073 \\
   & Norm & 0.0219297 \\
{\tt gabs} & Energy (keV) & 0.25 \\
   & Sigma (keV) & 0.153515 \\
   & Norm & 1543.75 \\
{\tt constant} & & 2.35765 \\
{\tt expdec} & Factor & 2.32895 \\
   & Norm & 0.0219297 \\
{\tt gaussian} & Energy (keV) & 0.451192 \\
   & Sigma (keV) & 0.0911681 \\
   & Norm & 0.0027839 \\
   \hline
   \end{tabular}

\begin{tablenotes}
\item[] Order of the models (top to bottom) is the same as that in text (left to right). Units of the ``Norm'' are photons keV$^{-1}$ s$^{-1}$ cm$^{-2}$ at 1~keV for {\tt powerlaw}, same at flat top for {\tt fsline}, same at 0~keV for {\tt expdec}, and total photons s$^{-1}$ cm$^{-2}$ for {\tt gaussian}.
\end{tablenotes}

\end{threeparttable}
\end{table}

\section{Comparison between the individual ACIS-stowed observation spectra and output spectra of {\tt mkacispback}}

\begin{figure}[htb!]
\centering
\includegraphics[width=16cm]{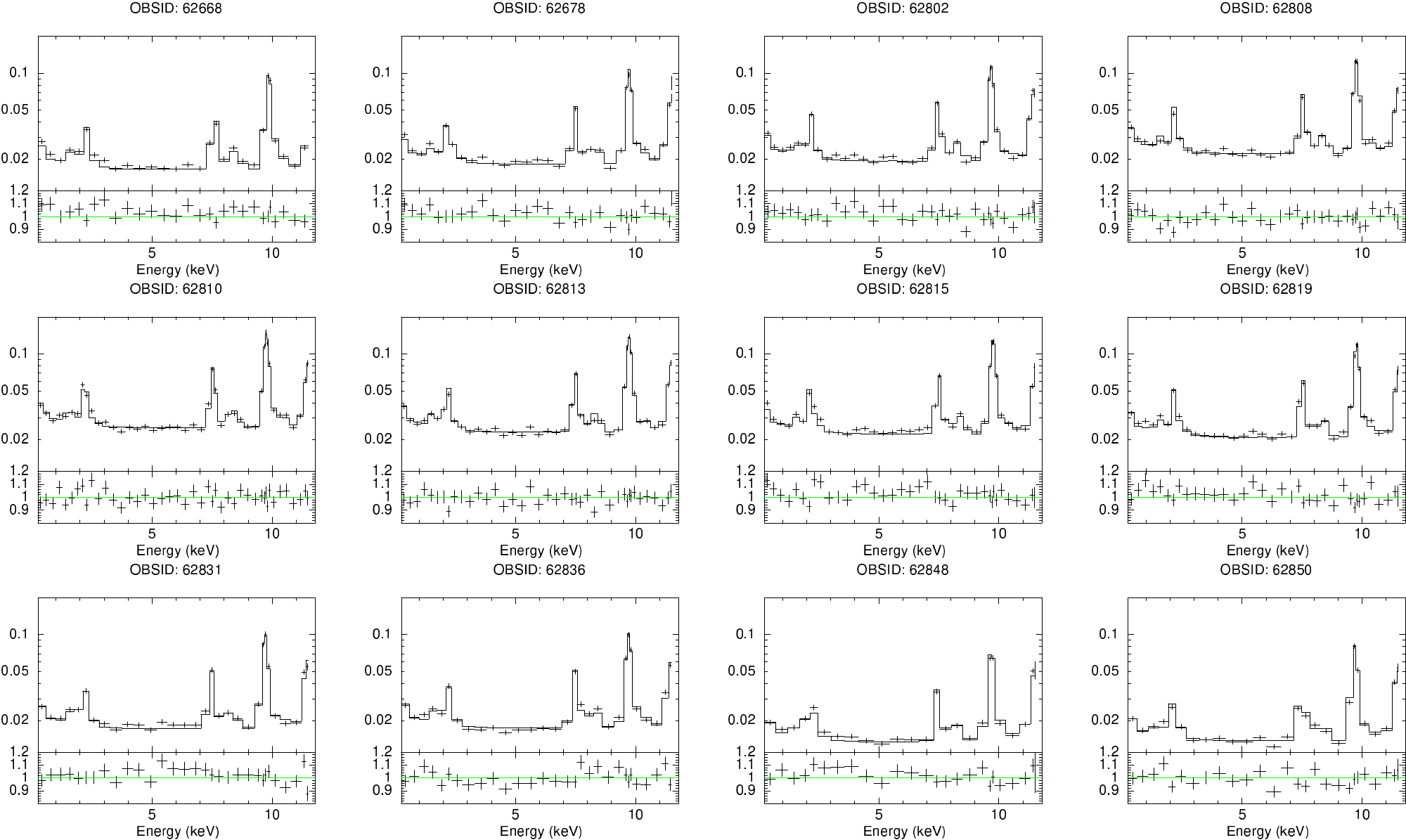}
\caption{Comparison between the individual ACIS-stowed observation spectra and output spectra of {\tt mkacispback} for ACIS-I0, VFAINT mode.
In each case, spectrum is extracted from the entire CCD.
In each panel, the upper and lower panel show the data and model count rates (s$^{-1}$ keV$^{-1}$) and their ratios (data/model), respectively.
One observation per year from the observation list (Table~\ref{tab-stowed}) is presented. Observations with exposure time of less than 40~ksec are omitted.
\label{fig-vf-i0-eachobs}}
\end{figure}


\begin{figure}[htb!]
\centering
\includegraphics[width=16cm]{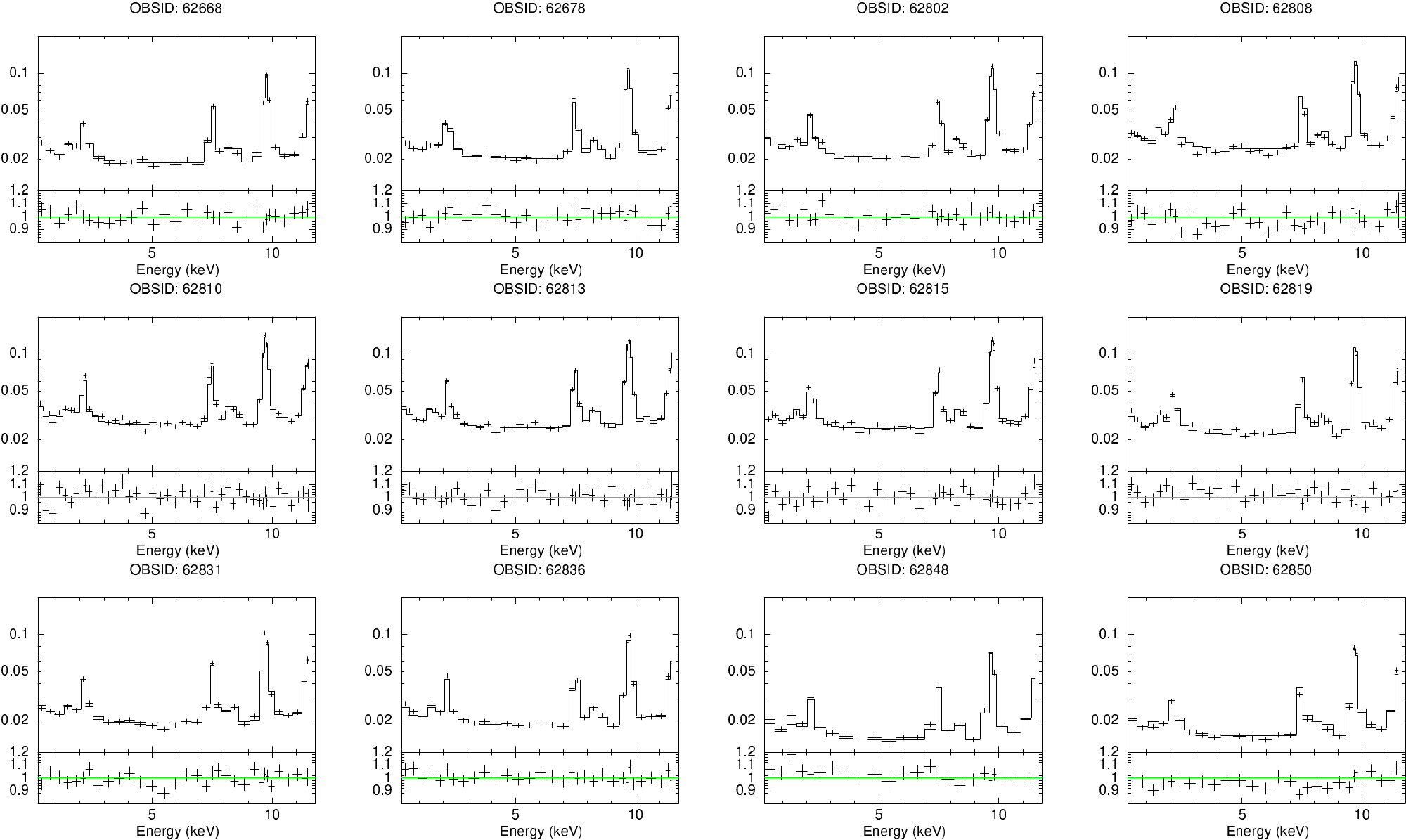}
\caption{Same as Fig.~\ref{fig-vf-i0-eachobs} but for ACIS-I3, VFAINT mode.
\label{fig-vf-i3-eachobs}}
\end{figure}

\begin{figure}[htb!]
\centering
\includegraphics[width=16cm]{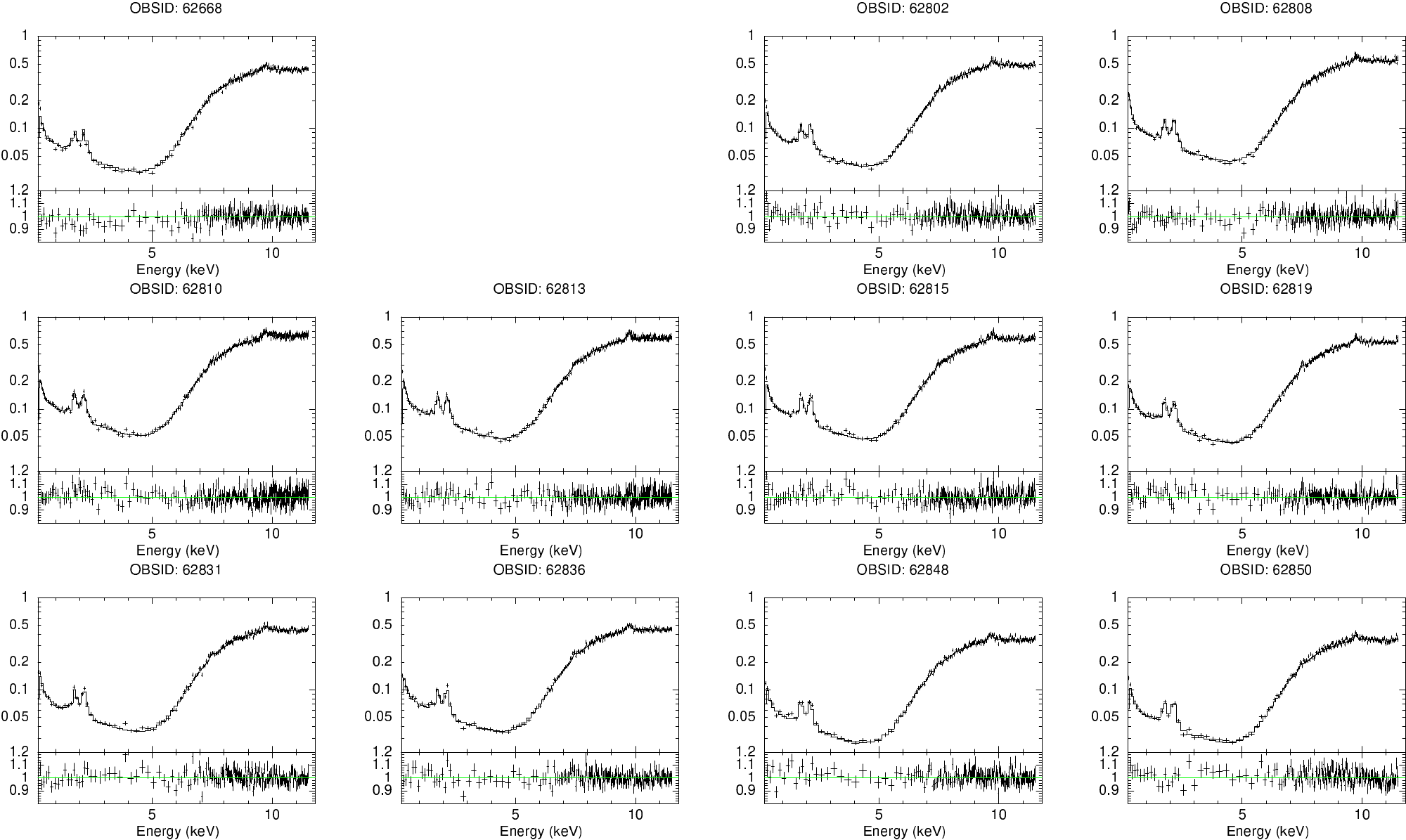}
\caption{Same as Fig.~\ref{fig-vf-i0-eachobs} but for ACIS-S1, VFAINT mode.
No S1 data are available for OBSID 62678; for that observation, ACIS-I1 was on instead of ACIS-S1.
\label{fig-vf-s1-eachobs}}
\end{figure}

\begin{figure}[htb!]
\centering
\includegraphics[width=16cm]{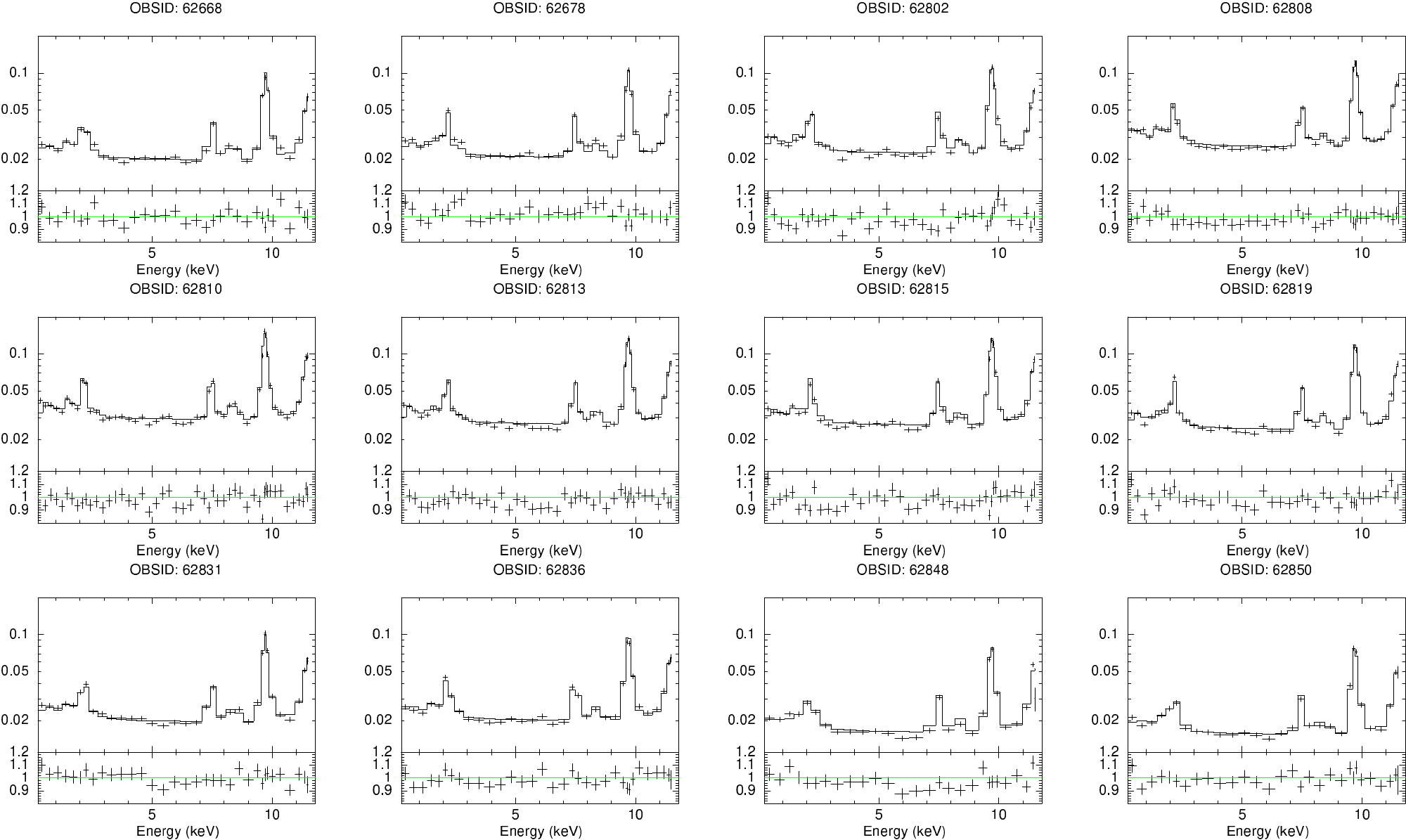}
\caption{Same as Fig.~\ref{fig-vf-i0-eachobs} but for ACIS-S2, VFAINT mode.
\label{fig-vf-s2-eachobs}}
\end{figure}

\begin{figure}[htb!]
\centering
\includegraphics[width=16cm]{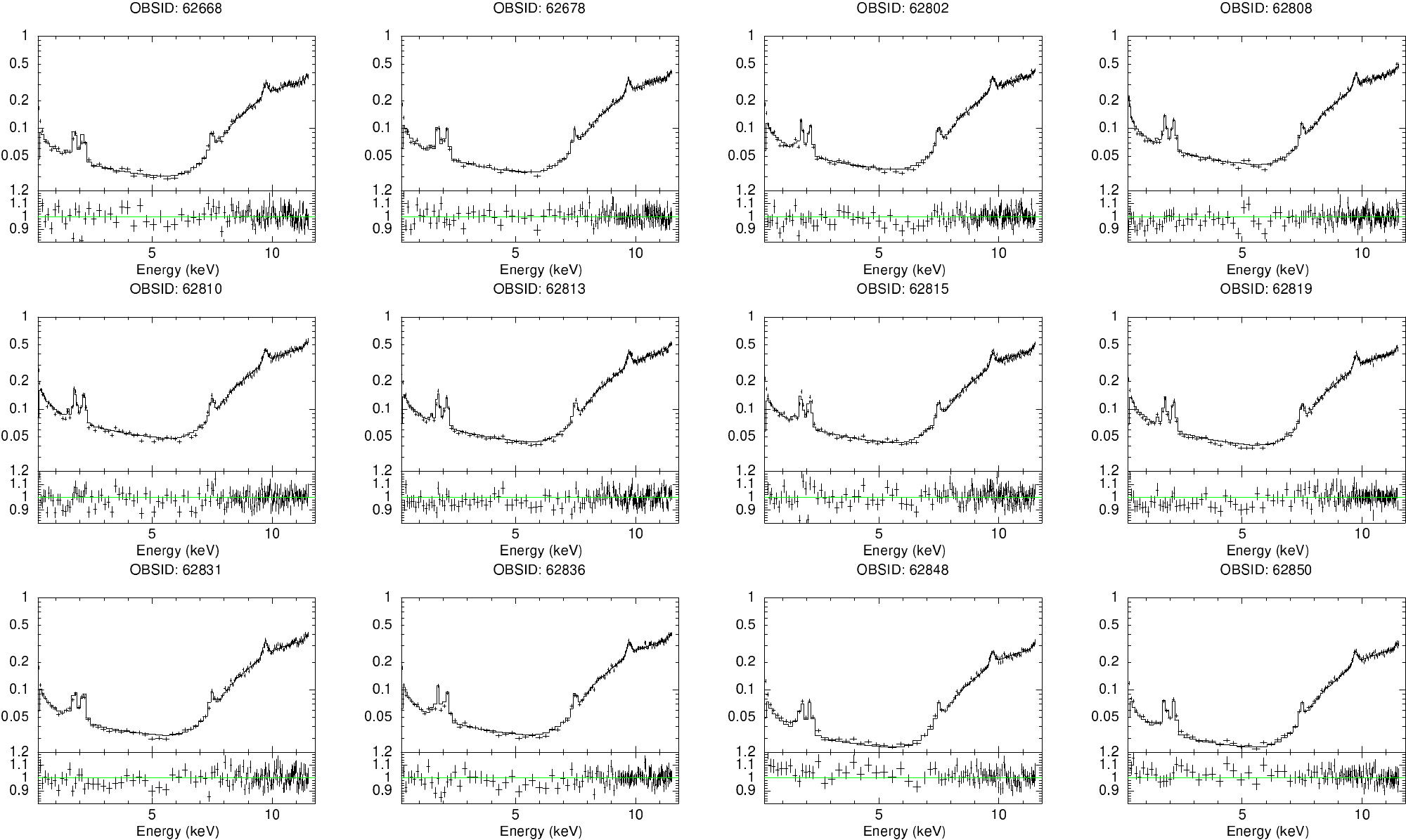}
\caption{Same as Fig.~\ref{fig-vf-i0-eachobs} but for ACIS-S3, VFAINT mode.
\label{fig-vf-s3-eachobs}}
\end{figure}

\begin{figure}[htb!]
\centering
\includegraphics[width=16cm]{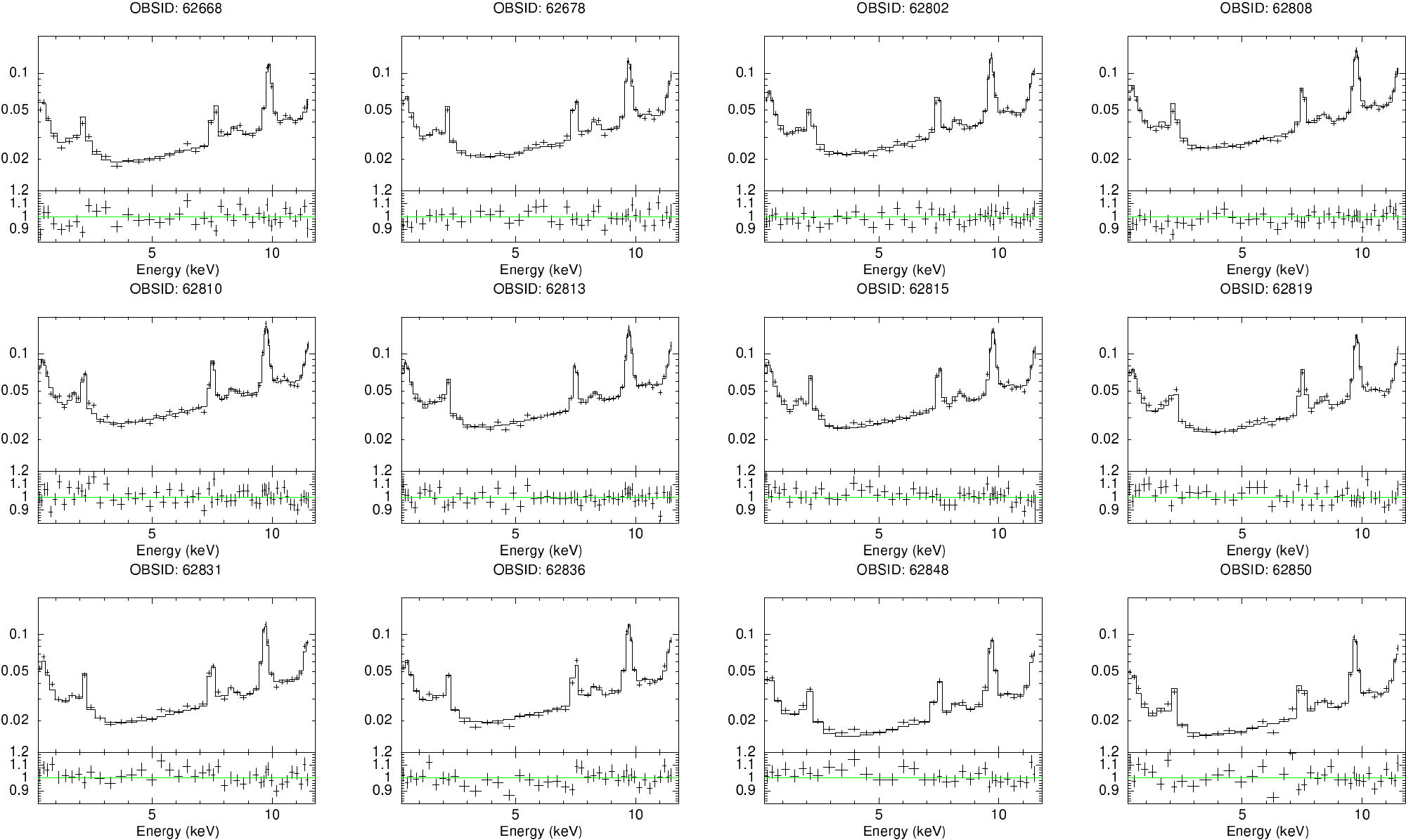}
\caption{Same as Fig.~\ref{fig-vf-i0-eachobs} but for ACIS-I0, FAINT mode.
\label{fig-f-i0-eachobs}}
\end{figure}


\begin{figure}[htb!]
\centering
\includegraphics[width=16cm]{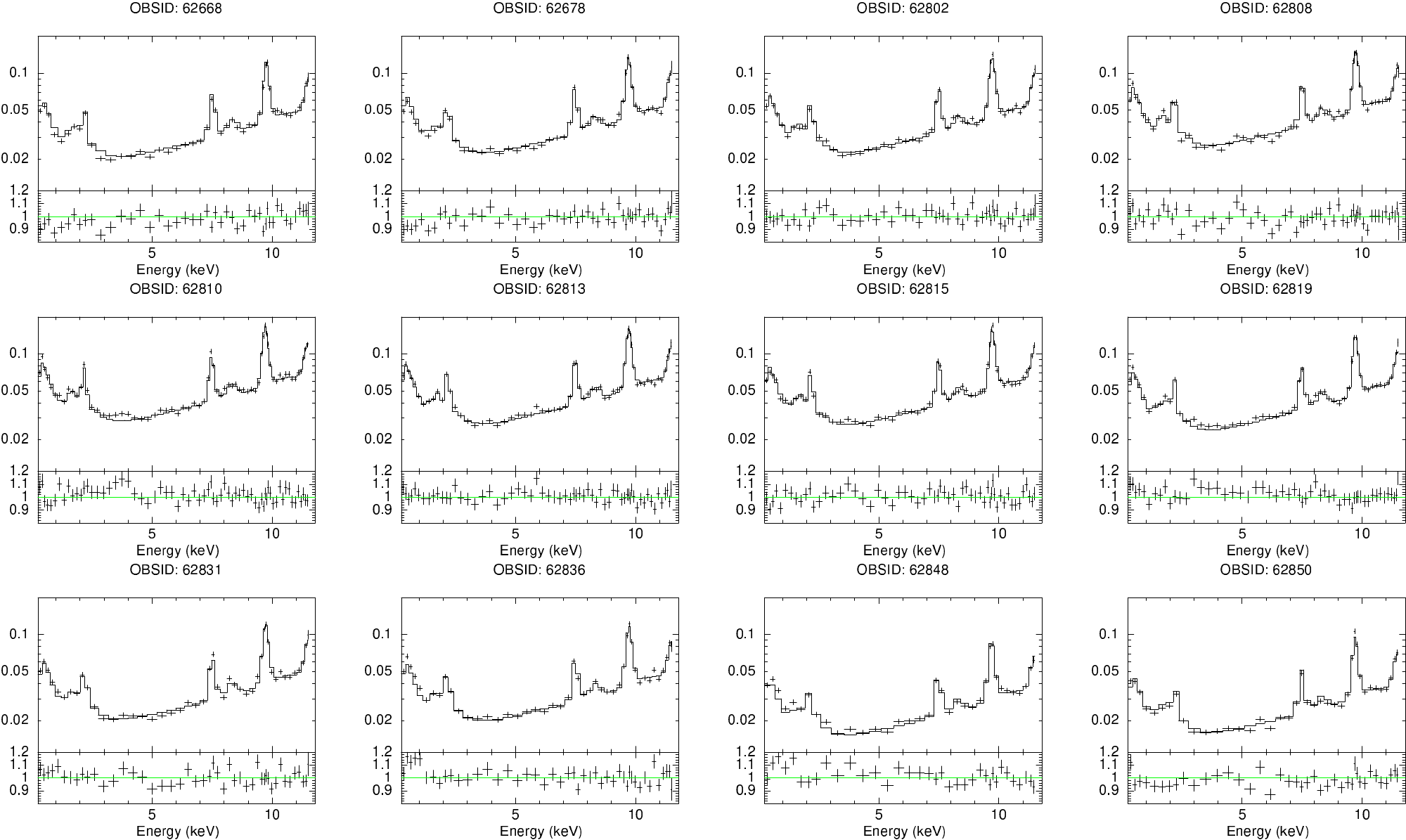}
\caption{Same as Fig.~\ref{fig-vf-i0-eachobs} but for ACIS-I3, FAINT mode.
\label{fig-f-i3-eachobs}}
\end{figure}

\begin{figure}[htb!]
\centering
\includegraphics[width=16cm]{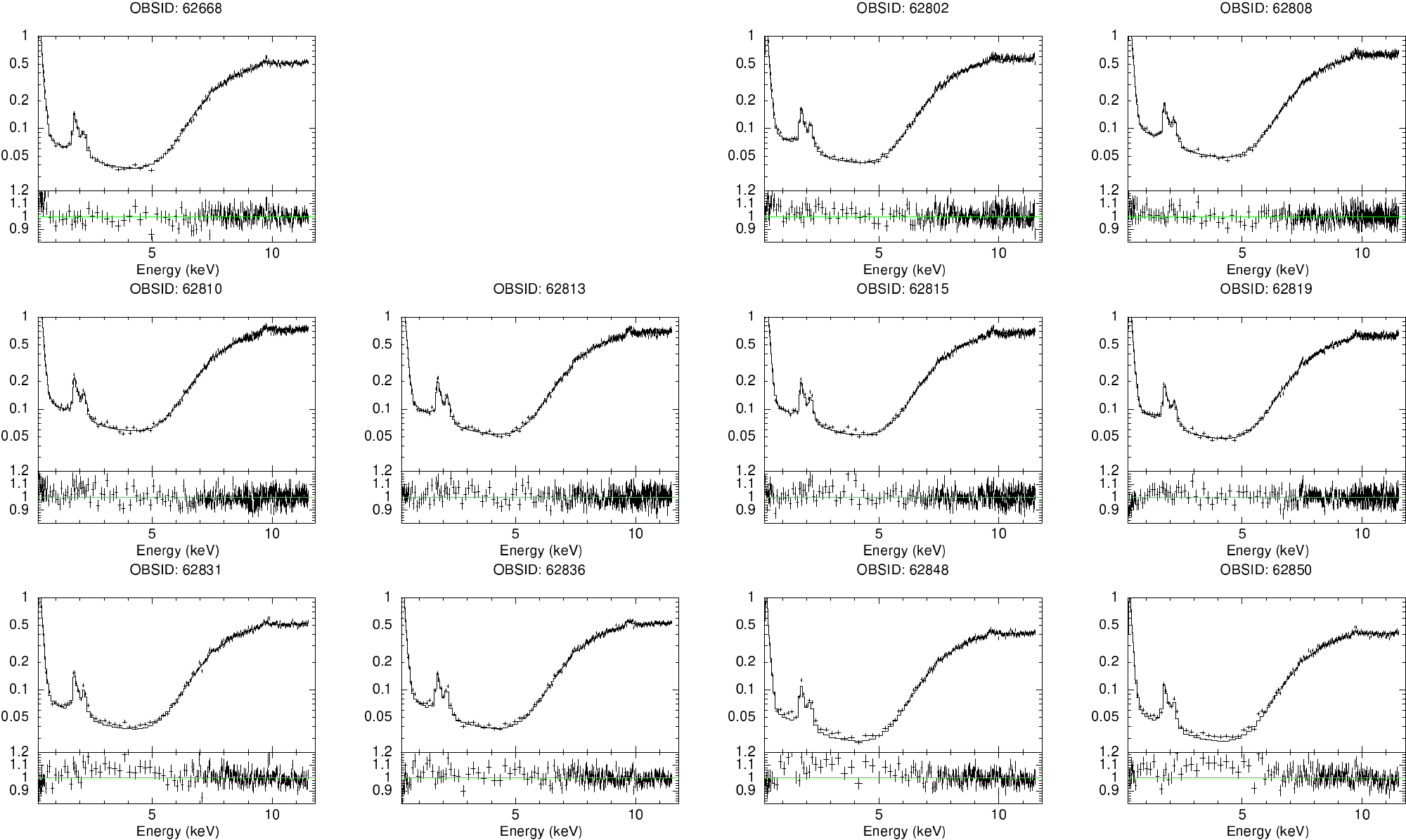}
\caption{Same as Fig.~\ref{fig-vf-i0-eachobs} but for ACIS-S1, FAINT mode.
No S1 data are available for OBSID 62678; for that observation, ACIS-I1 was on instead of ACIS-S1.
\label{fig-f-s1-eachobs}}
\end{figure}

\begin{figure}[htb!]
\centering
\includegraphics[width=16cm]{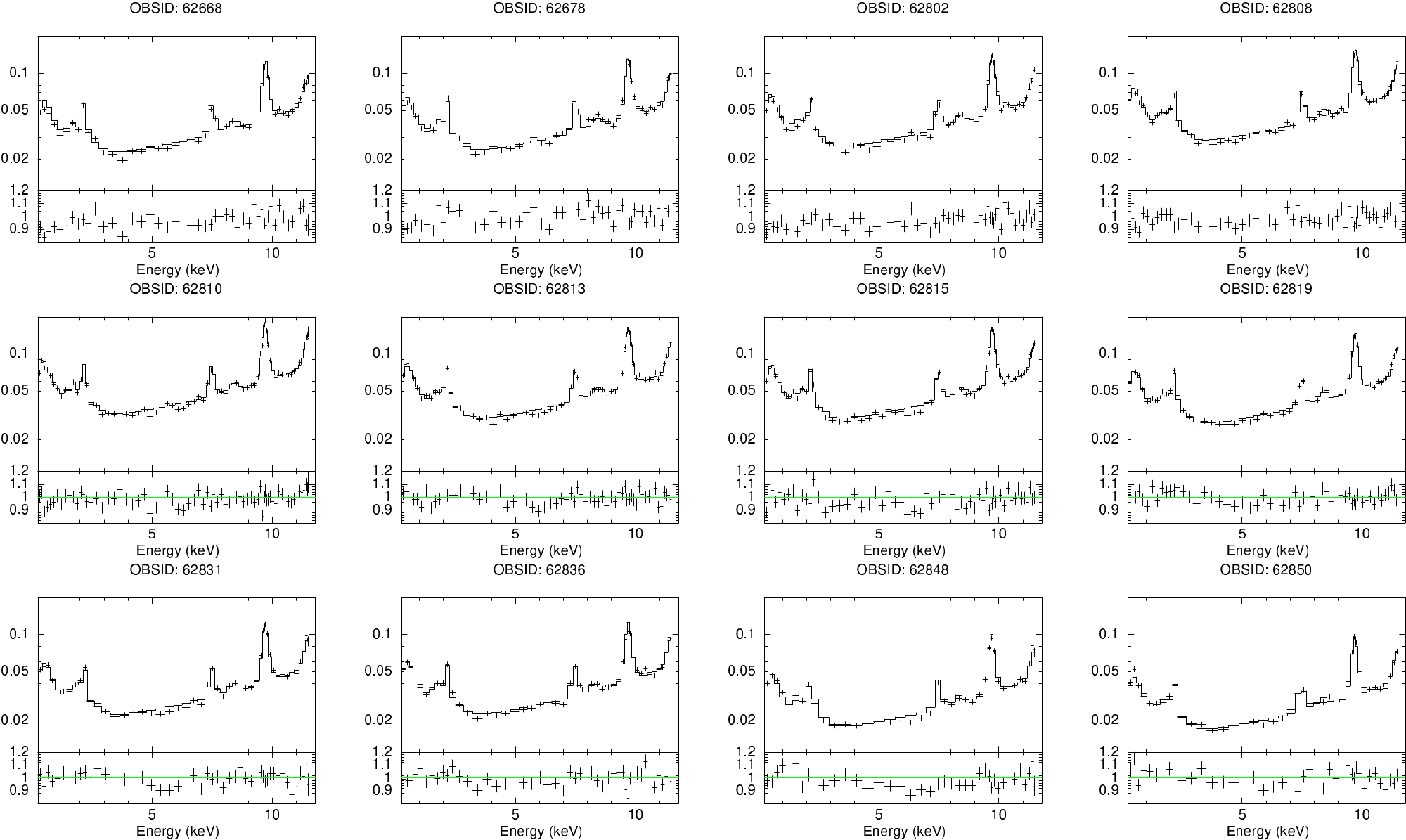}
\caption{Same as Fig.~\ref{fig-vf-i0-eachobs} but for ACIS-S2, FAINT mode.
\label{fig-f-s2-eachobs}}
\end{figure}

\begin{figure}[htb!]
\centering
\includegraphics[width=16cm]{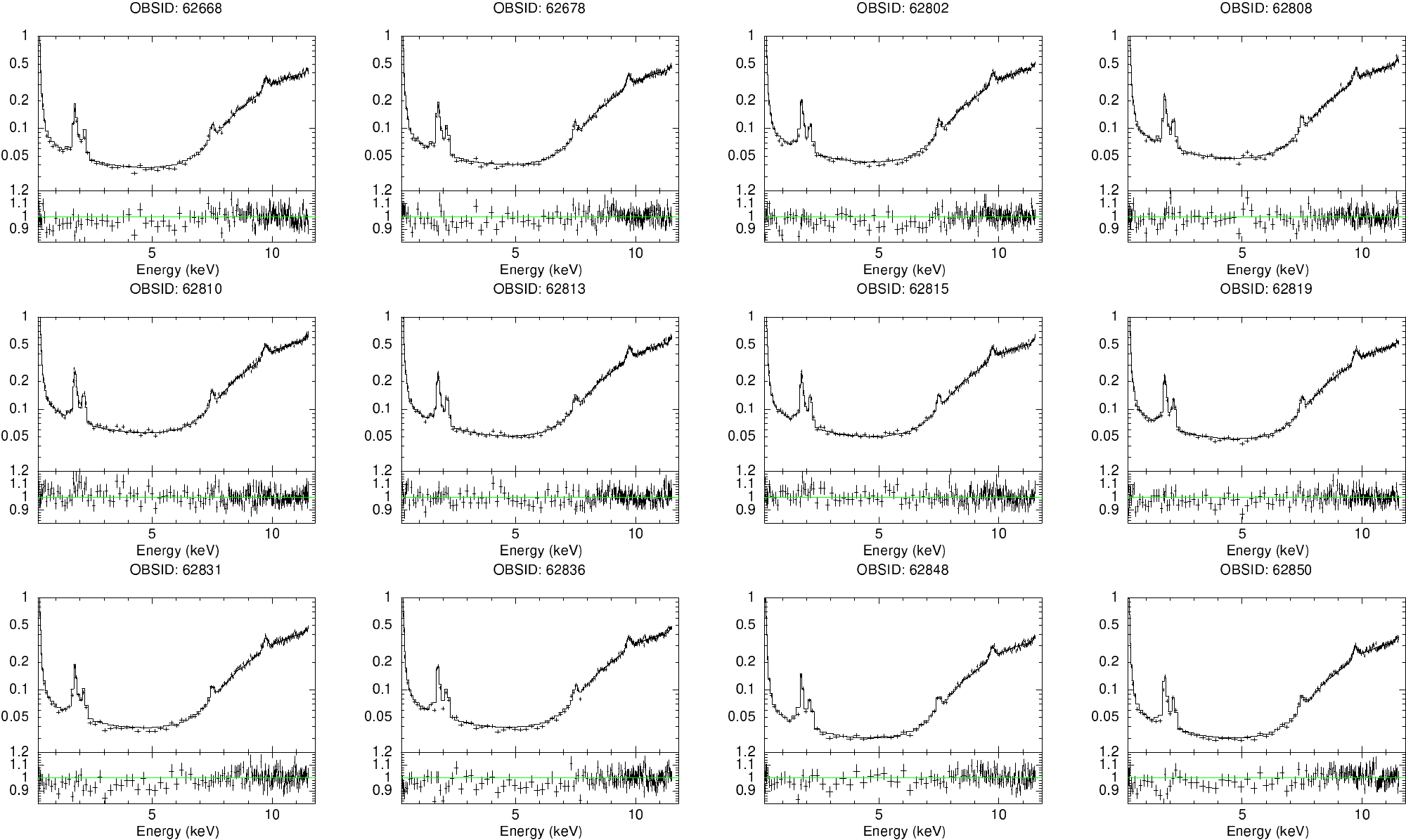}
\caption{Same as Fig.~\ref{fig-vf-i0-eachobs} but for ACIS-S3, FAINT mode.
\label{fig-f-s3-eachobs}}
\end{figure}

%
%

\end{document}